\documentclass[journal=jctcce,manuscript=article]{achemso}
\setkeys{acs}{doi = true} 

\usepackage[version=3]{mhchem} 
\usepackage{graphicx}
\usepackage{physics}
\usepackage{xcolor}
\usepackage{amsthm}
\usepackage{blkarray}
\usepackage{subcaption}
\usepackage{algorithm}
\usepackage{algpseudocode}
\usepackage{amsmath,amssymb}
\usepackage{hyperref}
\algrenewcommand\algorithmicrequire{\textbf{Input:}}
\algrenewcommand\algorithmicensure{\textbf{Output:}}

\author{Srushti Patil}
\affiliation[University of Copenhagen]
{NNF Quantum Computing Programme, Niels Bohr Institute, University of Copenhagen, 2100 Copenhagen, Denmark}
\author{Nina Glaser}
\affiliation[University of Copenhagen]
{NNF Quantum Computing Programme, Niels Bohr Institute, University of Copenhagen, 2100 Copenhagen, Denmark}
\alsoaffiliation[University of Copenhagen, Dept Chem]
{Department of Chemistry, University of Copenhagen, 2100 Copenhagen, Denmark}
\email{ngl@chem.ku.dk}

\title[An \textsf{achemso} demo]
 {Efficient targeting of arbitrary excited states with quantum inverse power iteration through filtering polynomials}

\abbreviations{IR,NMR,UV}

\begin{document}






\begin{abstract}
In this work, we introduce a quantum inverse power iteration (QIPI) algorithm based on the quantum singular value transformation (QSVT) to target arbitrary excited states.
Given an energy shift $\omega$, QIPI prepares the target excited state by iteratively applying an approximation of the shifted inverse Hamiltonian $(H-\omega I)^{-1}$ to a trial state.
Prior quantum inverse power approaches typically relied on Fourier decompositions of the inverse Hamiltonian, with numerical quadrature used to reconstruct the transformation, but such methods are highly sensitive to hyperparameter choices and have been observed to be numerically unstable, effectively restricting their use to ground-state preparation.
To enable robust excited-state targeting, we investigate two alternative transformation techniques: a Chebyshev decomposition of the inverse (Cheb-inv) and an eigenstate filtering (EF) approach based on QSVT.
We find that EF-based QIPI is substantially more robust than Cheb-inv and other decomposition-based approaches due to the symmetry of the applied filtering polynomial, avoiding divergence with respect to the choice of $\omega$ and efficiently suppressing off-target eigenstates even in closely spaced spectra.
Numerical simulations for molecular Hamiltonians of \ce{H2}, \ce{LiH}, and \ce{BeH2} show improved convergence and enhanced access to higher excited states relative to other quantum power methods.
Assuming standard oracle access to the Hamiltonian, we further provide logical resource estimates in fault-tolerant settings in terms of T gate counts, and conclude that QIPI can achieve high target state amplification with modest polynomial degrees, thereby making it a promising candidate for scalable excited-state preparation in fault-tolerant quantum chemistry applications.
\end{abstract}

\section{Introduction}

Characterizing molecular excited states plays a central role in numerous quantum chemical applications, including charge and energy transfer processes,\cite{charge_transfer} photochemical reactions,\cite{excited_photo} and the computation of molecular spectra.\cite{excited_states_appl, gonzalez2020quantum}
Despite their importance, the accurate determination of excited states remains a major challenge.
Unlike ground states, which can often be variationally targeted within a relatively compact region of Hilbert space,\cite{ground_state_variational_cc} excited states typically require access to a much larger configuration manifold to achieve comparable accuracy.\cite{large_hilbert_excited}

In practice, excited states are commonly computed using iterative and response-based techniques that target multiple eigenstates either sequentially or within state-averaged frameworks.
Widely used approaches include subspace-based diagonalization methods optimized for low-lying excitations,\cite{saad2011numerical} linear-response and equation-of-motion formalisms built upon ground-state references,\cite{Ullrich11_TDDFT,Krylov2008EquationofmotionCM} and multistate variational extensions that enforce orthogonality constraints or project out previously converged states.\cite{werner85_CASSCF,otis23_excitedstate}
Although these strategies avoid the full diagonalization of the Hamiltonian, their computational cost increases rapidly when targeting interior or higher-lying eigenstates, particularly in strongly correlated systems where the relevant configuration space expands substantially and memory requirements often become limiting.

\begin{figure*}
\centering
\includegraphics[width= 12cm, height=4cm]{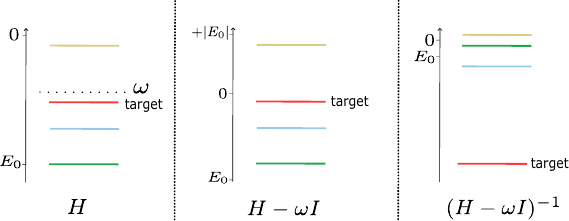}
\caption{Illustration of the spectral transformation resulting from shifting and inverting a given Hamiltonian $H$}
\label{fig:hamil_spectrum}
\end{figure*}

Another approach for eigenstates targeting is the use of iterative eigensolvers such as the power, inverse, and shift-and-invert methods to directly target specific states.\cite{saad2011numerical}
The power method illustrates the basic principle: repeated application of the Hamiltonian $H$ to a trial state amplifies the eigenstate component associated with the dominant eigenvalue $\lambda_1$, converging geometrically with the error decreasing as $\mathcal{O}\left(\left|\frac{\lambda_2}{\lambda_1}\right|^k\right)$ after $k$ iterations, with $\lambda_2$ being the first sub-dominant eigenvalue.
To access interior eigenstates, inverse power iteration (IPI) generalizes this idea by instead applying the shifted inverse operator $(H - \omega I)^{-1}$ with eigenvalues $\{(\lambda_i - \omega)^{-1} \mid i \in \mathbb{N}\}$. 
This shift-and-inversion makes the eigenvalue closest to $\omega$ dominant, thereby making the target eigenstate accessible through IPI (see Fig.~\ref{fig:hamil_spectrum} for an illustration).
Starting from an initial state $\ket{\Psi_0}$ having nonzero overlap with the target eigenstate with an eigenvalue $\lambda_t$, repeated application of $(H - \omega I)^{-1}$ yields geometric convergence with the error scaling as  $\mathcal{O}\left(\left|\frac{\lambda_t - \omega}{\lambda_j - \omega}\right|^k\right)$, where $\lambda_j$ is the next closest eigenvalue to $\omega$.
Despite these favorable convergence properties, direct numerical application of IPI to quantum many-body Hamiltonians is computationally prohibitive on classical hardware: both state vectors and operators grow exponentially with system size, making explicit matrix inversion intractable beyond modest system sizes.

Quantum computing offers a natural route to circumventing this exponential bottleneck by encoding and efficiently manipulating many-body wavefunctions in exponentially large Hilbert spaces through superposition and entanglement.\cite{feynman_simulating_1982, zalka1998simulating, kassal2011simulating, cao2019quantum}
To solve the eigenvalue problem associated with the time-independent Schrödinger equation, two paradigmatic approaches have emerged: quantum phase estimation (QPE)\cite{kitaev1995quantummeasurementsabelianstabilizer} and the variational quantum eigensolver (VQE).\cite{Peruzzo_2014}
QPE can achieve arbitrarily high precision in the retrieved energy and isolate the desired eigenstate with polynomial scaling, but requires deep coherent circuits and access to initial states with high overlap with the desired target state.\cite{ding2023even, lin2022heisenberg}
VQE replaces deep coherent evolution with a hybrid quantum--classical optimization loop suited to near-term devices,\cite{mccaskey2019quantum} and numerous extensions have been proposed to access excited states via penalty-based deflation, subspace expansions, spectral-shift strategies, and adaptive \textit{ansatz} constructions.\cite{imaginary_time_spectral, highly_excited_spectral_transforms, filter_enhanced_aqc,higgott2019variational, Nakanishi_2019, foldedspecvqe, Yordanov_2021}
However, these methods are typically most effective for low-lying states, require careful parameter tuning, and often exhibit slow or unstable convergence for higher excited states.
Quantum Krylov-based methods\cite{Krylov, stair2019multireferencequantumkrylovalgorithm,oliveira_quantum_2025} offer an intermediate alternative by constructing subspaces from time-evolved states with low-depth circuits, but similarly struggle to capture higher excited states reliably with standard Krylov subspace generators and face practical limitations from Trotter errors and high measurement costs.

As quantum hardware approaches early fault-tolerant regimes,\cite{2024BelowThreshold} there is an increasing need for algorithms that enable systematic targeting of interior eigenstates without requiring high-fidelity initial state preparation.
This motivates our quantum realization of the inverse power iteration algorithm, which requires no variational optimization, targets arbitrary eigenstates given only an energy estimate, and converges geometrically from any trial state with nonzero overlap with the target.
While previous quantum analogs of inverse power iteration have been proposed, most notably based on the Q-Inv framework,\cite{kyriienko2020quantum} these commonly rely on a Fourier decomposition of the inverse Hamiltonian implemented through linear-combination-of-unitaries (LCU).\cite{FourierHamSim}
These approaches are, however, highly sensitive to hyperparameter choices and have been observed to be numerically unstable, effectively restricting their practical use to ground-state preparation,\cite{cainelli2024numericalinvestigationquantuminverse} or reverting to hybrid quantum-classical schemes by relying on the variational quantum linear system solver.\cite{Yoshikura:2023ujd}

In this work, we present two different quantum inverse power iteration (QIPI) schemes, which approximate the action of $(H - \omega I)^{-1}$ through a) an alternative decomposition of the inverse with Chebyshev polynomials (Cheb-Inv) or b) a spectral transformation relying on eigenstate filtering (EF) polynomials.
Both schemes can be realized within the framework of quantum singular value transformation (QSVT).\cite{Gily_n_2019}
In the following, we first briefly summarize previous work on inverse power iteration quantum algorithms in Section~\ref{sec:litreview} before introducing our QIPI methodology in Section~\ref{sec:methods}.
We analyze the performance of our QIPI algorithm in Section~\ref{sec:results}, where we present both numerically exact simulations as well as results for QIPI applications in the fault-tolerant regime where arbitrary phase rotations are decomposed into finite gate sequences, and discuss resource requirements and the query complexity of the algorithm.
We further analyze QIPI with chemically informed, symmetry-adapted initial states, demonstrating improved convergence and characterizing the associated per-iteration and cumulative post-selection success probabilities.
We conclude our findings in Section~\ref{sec:conclusions} and outline future perspectives of QIPI for excited-states targeting in fault-tolerant quantum chemistry applications.

\section{Prior Inverse Power Iteration Quantum Algorithms}
\label{sec:litreview}

\begin{figure*}
\centering
\includegraphics[width= 8cm, height=4cm]{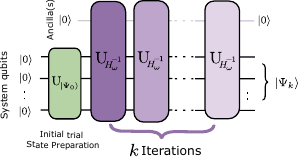}
\caption{Schematic illustration of QIPI. First, the system qubits are initialized and a trial state $\ket{\Psi_0}$ is prepared by applying a unitary $U_{\ket{\Psi_0}}$. 
Then, the unitary $U_{H_{\omega}^{-1}}$ which approximates the action of $H_{\omega}^{-1}=(H - \omega I)^{-1}$ is iteratively applied to obtain geometric convergence to the target eigenstate.
The unitary $U_{H_{\omega}^{-1}}$ could take different forms depending on the method used to express the inverse $H_{\omega}^{-1}$ as a unitary operator.
For instance, methods such as block encodings require additional ancillae, followed by post-selection in each iteration.}
\label{fig:gQIPI}
\end{figure*}

Power and inverse power methods have been extensively used in quantum chemistry to solve the eigenvalue problem and have recently been extended to the quantum domain.
To perform the iterative applications of the respective operator $O$ (where $O$ is usually the system Hamiltonian or a spectral transform thereof) on a quantum computer, one prepares an initial trial state $\ket{\Psi_0}$ via a state preparation unitary $U_{\ket{\Psi_0}}$ and then sequentially applies a unitary $U_{O}$ that approximates the action of the chosen operator, as illustrated in Figure~\ref{fig:gQIPI} for the quantum inverse power iteration.

The overarching challenge in power method-inspired quantum algorithms is thus how to encode the generally non-unitary Hamiltonian, or its inverse, into unitary operators that can be implemented on a quantum computer.
While there are several well-established techniques for implementing the effect of the Hamiltonian itself on a given quantum state, such as decomposing it into a linear combination of unitaries, the unitary encoding of the Hamiltonian inverse is commonly achieved by using Fourier or Chebyshev-decomposition approaches.\cite{Childs_2017} 
Specifically, in quantum power methods, Fourier decomposition schemes have been widely used to realize the inverse,
which admit a natural representation in terms of time-evolution unitaries.\cite{Childs_2017,kyriienko2020quantum,cainelli2024numericalinvestigationquantuminverse}
It is realized as follows: 
In classical approximation theory, the function $1/x$ admits integral representations such as
\begin{equation}
    x^{-1} = \int_0^\infty e^{-xy}\,dy
\end{equation}
for $x>0$, which can be discretized into sparse sums of exponentials.
In the quantum setting, analogous constructions allow one to approximate $H^{-1}$ as a linear combination of unitary time evolutions $e^{-iHt}$.
Building on this idea, a Fourier-based approach expresses the inverse as a double integral over unitary propagators,\cite{Childs_2017}
\begin{equation}
    H^{-1} = \frac{i}{\sqrt{2\pi}} \int_{0}^{\infty} dy \int_{-\infty}^{\infty} dz\, e^{-z^2/2} e^{-iyzH},
\end{equation}
which is discretized and implemented using Hamiltonian simulation as a subroutine.
Kyriienko extended this framework to higher inverse powers in the quantum inverse (Q-Inv) iterative algorithm,\cite{kyriienko2020quantum} constructing $H^{-k}$ as
\begin{equation}
    H^{-k} = \frac{iN_k}{\sqrt{2\pi}} \int_{0}^{\infty} dy \int_{-\infty}^{\infty} dz\, z y^{k-1} e^{-z^2/2} e^{-iyzH},
\end{equation}
where $N_k$ is a normalization factor.
Q-Inv demonstrates improved scaling in the number of iterations compared to the classical inverse power method,\cite{cainelli2024numericalinvestigationquantuminverse} but in its original formulation, primarily amplifies dominant eigencomponents and is therefore naturally suited to ground-state preparation.

Yoshikura \textit{et al.} attempted to extend the Q-Inv algorithm to excited states, but found that due to the gap $|\lambda_i - \omega|$ typically being very small for all but the ground state, long time evolutions are required to resolve excited states, thereby making Q-Inv challenging to apply to excited states in practice.\cite{Yoshikura:2023ujd}
As an alternative, they proposed an adaptive scheme that iteratively solves the underlying linear systems with a variational quantum linear system solver to prepare the inverse-power states.
While this methodology enables the direct targeting of excited states through an inverse power iteration-based scheme, it by default inherits the limitations common to variational quantum algorithms, such as their \textit{ansatz}-dependence, slow convergence due to Barren plateaus, and high measurement overheads.\cite{tilly_variational_2022}
More recently, the QSVT framework has already been applied to quantum power methods\cite{ImprovedQuantumPowerMethod}, but while the QSVT-based power method results in an improved scaling, its applicability remains restricted to finding only dominant eigenstates.
Another spectral transformation approach uses quantum signal processing to construct a polynomial that approximates the folded spectrum operator of the form $(H - \omega I)^2$. 
\cite{khinevich2025quantumpoweriteration}
While conceptually related to our QIPI methodology, 
the folded algorithm suffers from numerical instabilities: its polynomial construction cannot distinguish energy levels efficiently, as multiple eigenvalues in the folded spectrum can cluster together, greatly affecting the convergence of the method.

Taken together, existing quantum inverse iteration algorithms highlight the potential of spectral transformation techniques for eigenstate preparation, yet they are either tailored to dominant eigenstates, or suffer from poor convergence due to constructions used for approximating the action of the inverse, thereby motivating the development of alternative approaches for QIPI that enable the systematic targeting of interior eigenstates in fault-tolerant quantum computations.

\section{Methodology}
\label{sec:methods}
\subsection{Chebyshev Decomposition-based QIPI}\label{sec: Cheb-Inv IPI}
Complementary to the Fourier-based decomposition of the Hamiltonian inverse as summarized in Sec.~\ref{sec:litreview}, an alternative decomposition scheme based on Chebyshev polynomials has been previously proposed in the literature.\cite{Childs_2017}
In this approach, Chebyshev series were specifically employed within linear-system solvers, implemented through quantum walks, which require sparse-access oracles.
In our approach, however, we use the same polynomial approximation within a QSVT framework to approximate the inverse of the Hamiltonian, and hence, sparsity is not a necessary condition for the Chebyshev scheme in this work. 
With the Chebyshev approach, $x^{-1}$ is approximated as
\begin{equation}
x^{-1} \;\approx\; G_N(x)
\;=\;
4 \sum_{j=0}^{N} 
\underbrace{\left[\,
(-1)^j\sum_{i=j+1}^{b} \frac{1}{2^{2b}}\binom{2b}{\,b+i\,}
\right]}_{c_{2j+1}}
\mathcal{T}_{2j+1}(x),
\label{eq:cheb-inv}
\end{equation}
where $b = \kappa^2 \log (\kappa/\epsilon_c)$, with $\kappa$ being the condition number and $\epsilon_c$ the target accuracy. $\mathcal{T}_l(x)$ are degree-$l$ Chebyshev polynomials of the first kind, defined as $\mathcal{T}_0(x)=1$, $\mathcal{T}_1(x)=x$, and $\mathcal{T}_{l+1}(x)=2x\,\mathcal{T}_l(x)-\mathcal{T}_{l-1}(x)$.
Equivalently, they can be written as $\mathcal{T}_l(\cos\theta)=\cos(l\theta)$.
In the original construction,\cite{Childs_2017} the truncation order $N$ is chosen as a function of $\kappa$ and the target accuracy $\epsilon_c$, which also determines the maximum polynomial degree $N_{\text{max}} = 2N+1$.
As high-order polynomials can suffer from numerical instabilities when determining the expansion coefficients $c_{2j+1}$, while also resulting in very high resource requirements for gate-based implementations, we examine lower-order truncations as an empirical parameter to probe the practical applicability of Chebyshev-based QIPI approaches to chemical problems.

To realize our Chebyshev-based QIPI algorithm, we first shift the Hamiltonian with the energy guess $\omega$ of the target state, setting $H_{\omega} = H - \omega I$ and define the scaled operator $\tilde{H} = H_{\omega}/s$, where the scaling parameter $s = \max (\abs{\lambda_{min} - \omega}, \abs{\lambda_{max} - \omega})$ ensures that the spectrum of $\tilde{H}$ is bounded by $[-1,1]$. 
If $\mu = \min \abs{\tilde{\lambda}_i}$ and $\kappa = 1/\mu$, then the spectrum of $\tilde{H}$ lies in $\mathcal{D}_\kappa$ = $[-1,-1/\kappa]\cup[1/\kappa,1]$.
Because $\mathcal{D}_{\kappa}$ excludes the origin, where $f(x)=1/x$ diverges, the inverse function can be approximated on this domain. 
Consequently, since $\tilde{H}$ is Hermitian, this polynomial approximation to $f(x)$ directly yields an approximation to $\tilde{H}^{-1}$.
We therefore consider the Chebyshev expansion $G_N(x)=\sum_{j} c_{2j+1}\,\mathcal{T}_{2j+1}(x)$, with coefficients as given in  Eq.~\eqref{eq:cheb-inv} to approximate the inverse of $\tilde{H}$.

\begin{figure}[t]
  \centering
  \includegraphics[width= 15cm, height=7cm]{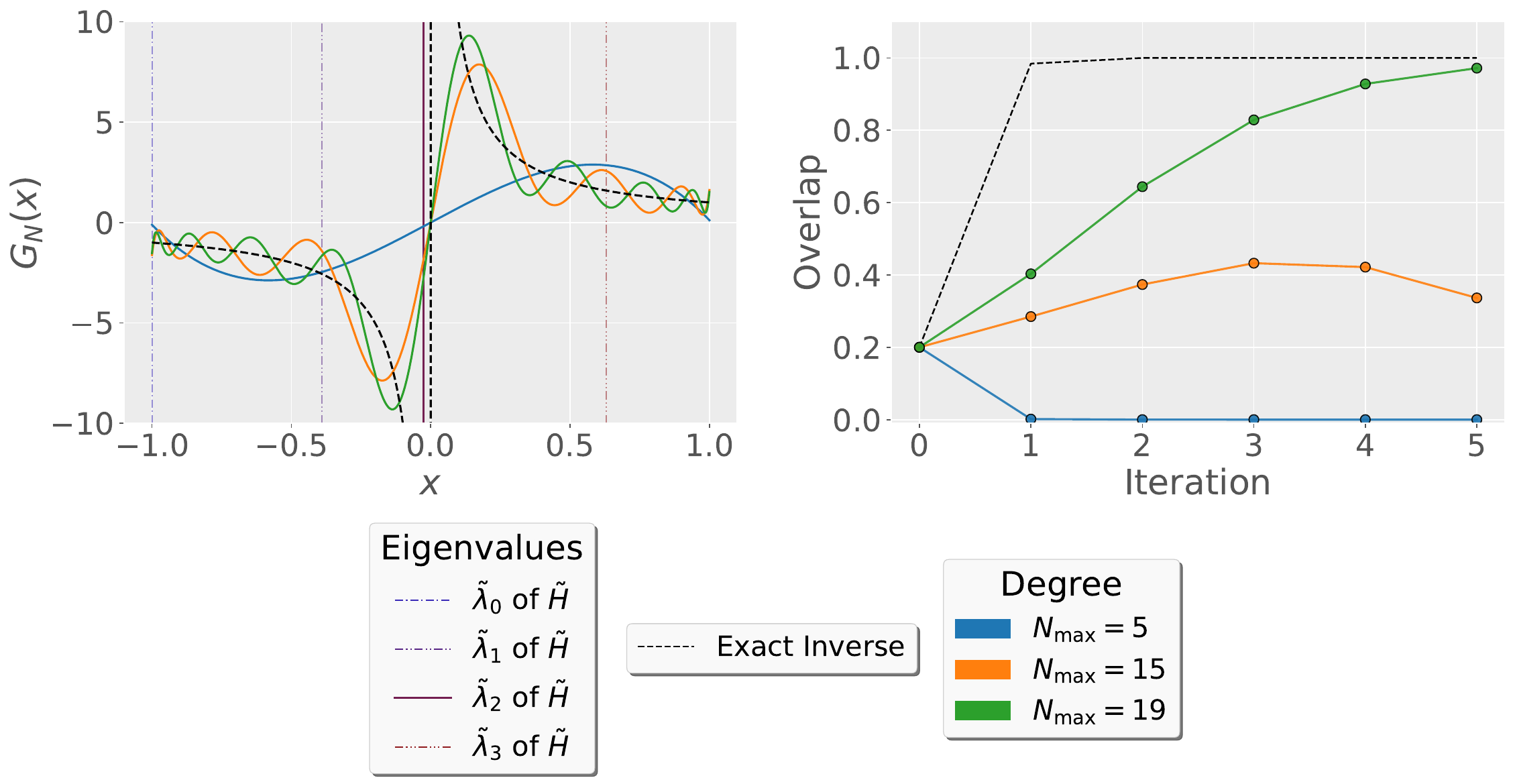}
  
  \caption{Cheb-Inv QIPI application to \ce{H2} in the STO-3G basis.
Left: The inverse of the shifted and scaled Hamiltonian $A$ is decomposed with Chebyshev polynomials with different truncation orders to target the eigenstate corresponding to $\lambda_2$.
Right: Chebyshev-based QIPI convergence as measured by the target state fidelity of the projected trial state for different maximum polynomial degrees $N_{\text{max}}$.
The black dashed line corresponds to the application of the exact inverse.}
  \label{fig:chebdecomp}
\end{figure}

We provide an illustrative application of the Chebyshev-based QIPI approach in Fig.~\ref{fig:chebdecomp} where we numerically simulate its performance for molecular hydrogen in the STO-3G basis.\cite{sto-3g}
We choose a shift $\omega$ such that the second excited eigenvalue $\lambda_2$ is the one closest to the shift, and therefore the corresponding shifted eigenvalue $\tilde{\lambda}_2$ has the smallest absolute value. 
Repeated application of $G_N(\tilde{H})$ amplifies the target component only when the polynomial response at $\tilde{\lambda}_2$ dominates compared to the response of all remaining eigenvalues. 
For sufficiently large polynomial degrees, this condition is satisfied, and the iteration converges to the target state. 
For smaller degrees, however, the approximation does not reproduce the sharp growth of $1/x$ near the excluded interval around the origin, so off-target eigencomponents may be amplified at a comparable or larger rate, leading to convergence to the wrong state.
Compared to an exact application of the inverse, Cheb-Inv QIPI requires more iterations to reach convergence, as expected due to the reduced amplification potential of the approximation as determined by the maximum peak height of the chosen polynomials.
When looking at the lower-degree polynomial approximations, it becomes apparent that the extrema of these decompositions are located far from the origin.
Thus, the function response at the target eigenenergy, which is closest to the origin of the shifted spectrum by default, is low, while other eigenstates are amplified in comparison.
Consequently, Cheb-Inv QIPI fails to converge for lower polynomial degrees, a behavior which is exacerbated further as the quality of the energy guess is improved, since then $|\tilde{\lambda}_t - \omega|$ decreases.
In that regime, $\mu$ becomes small, the effective condition number $\kappa$ increases, and the polynomial degree required for a faithful approximation grows rapidly, whereas for the exact inverse, the convergence would be accelerated as $|\tilde{\lambda}_t - \omega|\rightarrow 0$.
It can also happen that, during the initial iterations, Chebyshev-based QIPI appears to converge, for example for $N_{\max}=15$ (orange curve in Fig.~\ref{fig:chebdecomp}), but eventually amplifies the wrong state. 
This pseudo-convergence can arise because the Chebyshev polynomial decompositions of the inverse for smaller degrees do not decay monotonically away from zero; as a result, off-target components in the initial state may be amplified, causing convergence toward an incorrect state.
Therefore, we find Chebyshev-decomposition-based schemes to be unsuitable for targeting excited states, as they require very high polynomial degrees to target interior states embedded in dense spectra, and since they possess the undesirable property that the required polynomial degree increases rather than decreases with the quality of the energy guess.
Thus, these approximations require high computational cost while remaining unstable with respect to the energy shift, as near-exact guesses will lead to failure of the method at practically implementable polynomial degrees, which in practice makes the method unsuited to target any eigenstate other than the ground state.
These shortcomings motivate the development of more stable constructions of polynomial decompositions that remain well-conditioned near the spectral origin, thus improving convergence with better initial guesses, and extending also to Hamiltonians that contain more densely spaced eigenstates.
A promising alternative to Cheb-Inv QIPI is an iterative spectral filtering scheme through eigenstate filtering polynomials, which offer a particularly stable and efficient approach for targeting interior eigenstates, as introduced in the following.

\subsection{Eigenstate Filtering Polynomial} 
As before, let $\tilde{H}$ be the shifted and scaled operator $\tilde{H} = \frac{H - \omega I}{s}$ and consequently $\tilde{\lambda}_i = \frac{\lambda_i - \omega}{s}$,
so that the spectrum is spaced maximally between $-1$ and $1$. 
Suppose the target eigenvalue $\lambda_t$ is the unique one closest to the shift, meaning there exists $\Delta > 0$ such that $\abs{\tilde{\lambda}_t} < \Delta$ and $\abs{\tilde{\lambda}_i} \ge \Delta$ for all $i \neq t$. 
Our goal is to retain the $\lambda_t$–eigenstate while suppressing all other eigenstates.
If we consider $P_{\lambda_t}$ as a projector onto the $\lambda_t$–eigenspace, the underlying idea of eigenstate filtering is to construct a polynomial $P$ with $P(0)=1$ and with $|P(x)|$ small on the remaining spectrum $\mathcal D_{\Delta}:= [-1,-\Delta]\cup[\Delta,1]$, such that:
\begin{equation}
\label{eq: P_A}
P(\tilde{H}) = P\!\left(\frac{H-\omega I}{s}\right)\;\approx\; P_{\lambda_t}.
\end{equation}

\begin{figure}[t]
  \centering
     \includegraphics[width=0.7\columnwidth]{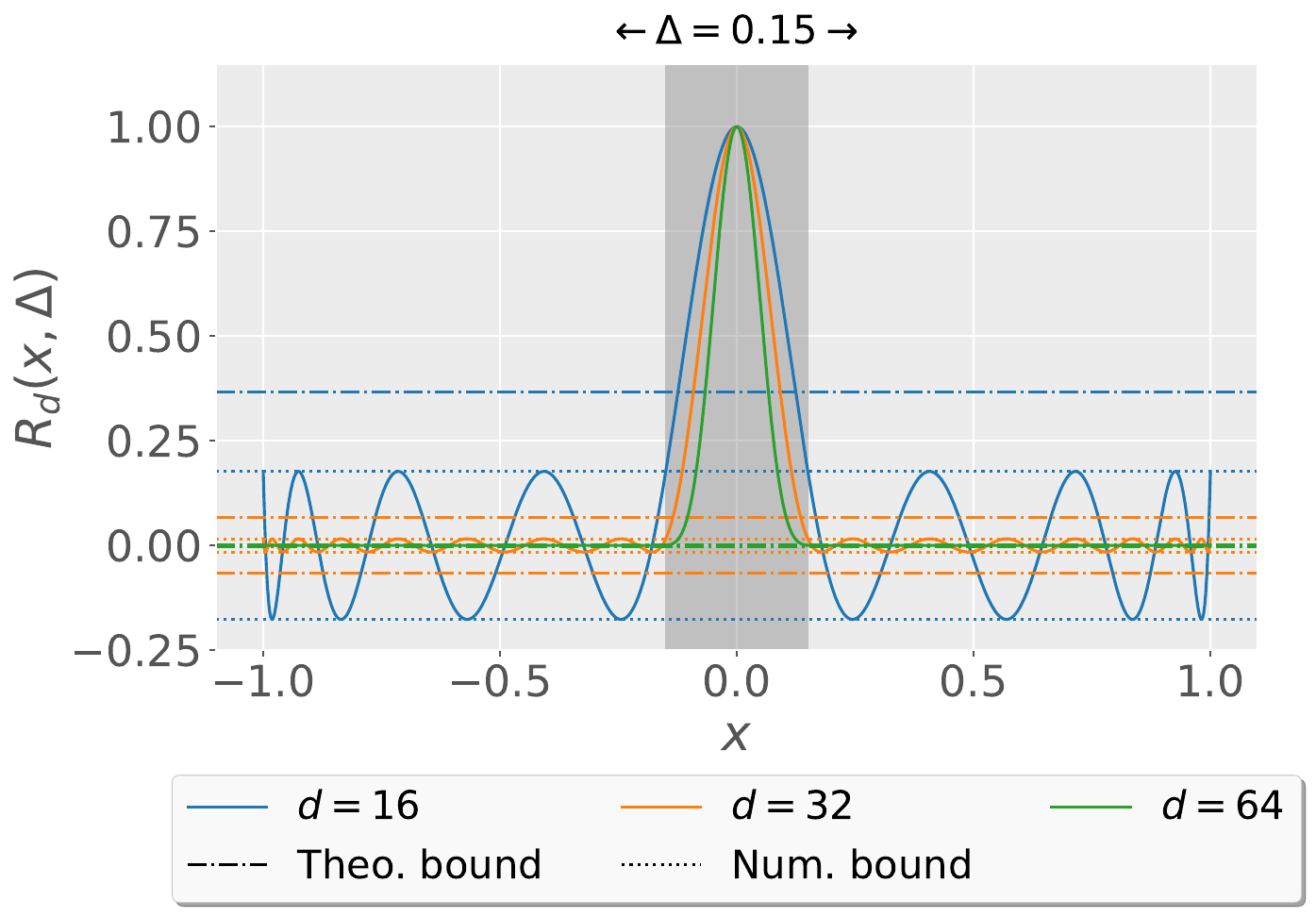}
  \caption{Eigenstate filtering polynomials with different degrees $d$ for $\Delta = 0.15$ (solid lines), with worst-case analytic bounds given by $2 e^{-\sqrt{2}\ell \Delta}$ (dash-dotted lines) and numerical bounds given by $\max_{x \in \mathcal D_{\Delta}}R_d(x, \Delta)$ (dotted lines) for a given degree.}
  \label{fig:minmax}
\end{figure}

A particularly suitable choice for $P$ is the $d = 2\ell$-degree eigenstate filtering polynomial (EFP) proposed by Lin and Tong\cite{Lin_2020}:
\begin{equation}
R_d(x;{\Delta})
\;=\;
\frac{ \mathcal{T}_\ell\!\left(-1 + \dfrac{2(x^{2}-{\Delta}^{2})}{\,1-{\Delta}^{2}\,}\right) }
     { \mathcal{T}_\ell\!\left(-1 + \dfrac{2(-{\Delta}^{2})}{\,1-{\Delta}^{2}\,}\right) },
\label{eq:minimax-filter}
\end{equation}
where $\mathcal{T}_\ell$ is the $\ell$-th Chebyshev polynomial of the first kind.  
This construction was inspired by a shifted-and-rescaled Chebyshev approximation and yields a polynomial that is maximal at $x=0$ while remaining small in $\mathcal D_{\Delta}$.
It can be shown that the EFP as given in Eq.~\eqref{eq:minimax-filter} is optimal for eigenstate filtering, as it yields the largest compression of unwanted components among all real polynomials satisfying $P(0) = 1$ up to degree $d$.\cite{Lin_2020}
For any window $0<\Delta \leq 1/\sqrt{12}$, the EFP is theoretically bounded by $|R_d(x;{\Delta})| \leq 2 e^{-\sqrt{2}\ell \Delta}$ in $ \mathcal D_{\Delta}$, corresponding to a exponential suppression of off-target spectral components.
As can be seen in Fig.~\ref{fig:minmax}, this worst case theoretical bound is loose such that even better compression can be achieved in practice as shown by the numerical bound given by $\max_{x \in \mathcal D_{\Delta}}R_d(x, \Delta)$.
To implement the eigenstate filtering polynomials in fault-tolerant quantum computing applications, we utilize the quantum singular value transformation framework.

\subsection{Quantum Singular Value Transformation}
\label{subsec:qsvt}

We here briefly present the QSVT framework, which enables polynomial transformations of matrices on a quantum computer, before introducing our EFP-based QIPI algorithm.
Since QSVT builds on quantum signal processing (QSP), we defer a more general introduction to QSP to Appendix~\ref{app:qsp}.

Any matrix $A$ of size ${n \times n}$ can be written in terms of its singular value decomposition as
\begin{equation}
    A = W_{\Sigma}\Sigma V^{\dagger}_\Sigma,
\end{equation}
where $W_\Sigma$ and $V_\Sigma$ are unitary matrices and $\Sigma$ is a diagonal matrix containing the singular values $\{\sigma_i\}_{i=1}^n$. Columns of $W_\Sigma$ and $V_\Sigma$ form orthonormal bases, which we denote by $\{\ket{w_k}\}$ and $\{\ket{v_k}\}$, respectively. 
Here $\{\ket{w_k}\}$ and $\{\ket{v_k}\}$ are the left and right singular vectors (spanning the left and right singular subspaces), respectively. 
With this notation, the singular value decomposition of $A$ can be rewritten as:
\begin{equation}
A \;=\; \sum_{k=1}^{n} \sigma_k \ket{w_k}\!\bra{v_k}.
\label{eq:svd-rank-r}
\end{equation}

\begin{figure}[t]
  \centering
     \includegraphics[width= 15.5cm, height=4cm]{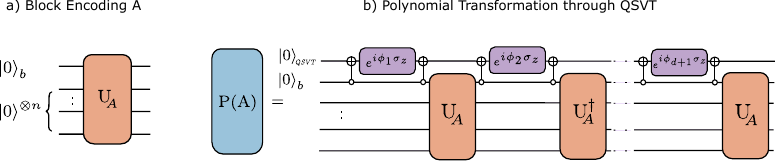}
  \caption{Schematic of QSVT. (a) First, we block encode our matrix $A$ into a larger unitary $U_A$ by using a dilation-based encoding with an extra ancilla $\ket{0}_b$ (b). The desired polynomial transformation is realized by alternate applications of the block encoding unitary and phase-controlled projectors.}
  \label{fig:qsvt}
\end{figure}

For a given polynomial $P(x)$, such as the EFP in Eq.~\eqref{eq:minimax-filter}, we want to perform a polynomial transformation of the form $P(A)$ to a given matrix $A$ on a quantum device. 
In our application, this matrix is the shifted and rescaled Hamiltonian $\tilde{H}$ and we want to construct a quantum circuit that realizes Eq.~\eqref{eq: P_A}.
To realize this transformation, we first \textit{block encode} $A$ into a larger unitary $U_A$ as
\begin{equation}
U_A \;=\;
\begin{blockarray}{c cc}
   & \Pi &   \\
\begin{block}{c[cc]}
\tilde{\Pi} & \;A & *  \;\; \\
 & \;*  \;\; & *  \;\;\\
\end{block}
\end{blockarray},
 \; \;A \;=\; \tilde{\Pi}\, U_A\, \Pi,
\label{eq:block-encoding}
\end{equation}
where $\tilde{\Pi}:=\sum_{k}\ket{w_k}\!\bra{w_k}$ and $\Pi:=\sum_{k}\ket{v_k}\!\bra{v_k}$ are projectors that locate $A$ inside $U_A$. 
In this particular case, $A$ is located in the upper left corner of $U_A$ and can be accessed by the projectors of the form: $\Pi = \tilde{\Pi} = (\ket{0_b}\bra{0_b})$.
Intuitively, $\Pi$ selects the columns and $\tilde{\Pi}$ selects the rows in which $A$ is embedded. 
While a multitude of different block encodings, such as LCU or QROAM with sophisticated SELECT and PREP operators (and additional ancilla overhead) with different QSVT conventions can be achieved,\cite{martyn2021grand} in this work, for demonstration purposes, we chose a dilation-based block encoding which is easy to construct and embeds $A$ into $U_A$ as: 
\begin{equation}
U_A \;=\; \raisebox{+.2\height}{$
\begin{blockarray}{c cc}
   & 0 & 1 \\
\begin{block}{c[cc]}
0 & A & i\sqrt{I-A^{2}}\\
1 & -i\sqrt{I-A^{2}} & A\\
\end{block}
\end{blockarray}
$}.
\label{eq: qipi_block_en}
\end{equation}
Here, 0 and 1 are labels for the sub-blocks in $U_A$, and we can easily verify that indeed, $U_A$ is a unitary matrix when $A$ is a Hermitian contraction with $\norm{A} \leq 1$. Such a block encoding can be realized in a quantum circuit with a single ancilla qubit as shown in Fig.~\ref{fig:qsvt} (a).

Given such a block encoding $U_A$, it is possible to exactly apply polynomial transformations on the target matrix $A$ through the QSVT framework.
The transformation is realized by alternately applying a phase-controlled version $\Pi_\phi$ and $\tilde{\Pi}_\phi$ of the projectors $\Pi$ and $\tilde{\Pi}$ respectively, as shown in Fig.~\ref{fig:qsvt} (b).
Mathematically they are phase-modulated reflections of the form $e^{i\phi (2\Pi - I)}$ and $e^{i\phi (2\tilde{\Pi} - I)}$,
which are simply $R_Z(\phi) = e^{i\phi\sigma_z}$ rotations sandwiched between controlled-NOT (CNOT) gates.
Thus, in addition to the block encoding ancilla $\ket{0}_b$, one extra ancilla $\ket{0}_{\text{QSVT}}$ is needed for the control of the QSVT rotation operations.

For an even degree-$d$ polynomial, such as the EFP one, the QSVT operator can then be written as
\begin{equation}
U_A^{\vec{\phi}}
\;=\;
\Bigg[\,
\prod_{k=1}^{d/2}
\Pi_{\phi_{2k-1}}\,U_A^{\dagger}\,\tilde{\Pi}_{\phi_{2k}}\,U_A
\Bigg]\Pi_{\phi_{d + 1}}
\;=\;
\begin{blockarray}{c cc}
   & \Pi &   \\
\begin{block}{c[cc]}
\tilde{\Pi} & P(A) & \cdot\;\;\\
 & \;\cdot  \;\; & \cdot  \;\;\\
\end{block}
\end{blockarray},
\label{eq:qsvt-even}
\end{equation}
where $P(A)$ can be obtained as $P(A)
\;:=\;
\sum_{k}\,
P(\sigma_k)\,\ket{v_k}\!\bra{v_k}.$
This sequence of operators thus implements the prescribed polynomial transformation on the embedded matrix $A$ by transforming its singular values.
For the matrices considered in this work, i.e. molecular Hamiltonians, this is equivalent to a polynomial transformation of their eigenvalues.

The rotation angles $\vec{\phi} = (\phi_1, .., \phi_{d + 1})$ needed to implement a specific polynomial transformation only depend on the functional form of the polynomial itself, and are independent of the input state or the Hamiltonian, and thus need only be pre-computed once for a given EFP.
For the numerical simulations presented in this work, we obtained the required phase angles with the \texttt{nlft-qsp} package.\cite{laneve2025generalizedquantumsignalprocessing, laneve2025nlftqsp}
To ensure the numerical stability of the phase angle computation in the preprocessing step of the \texttt{nlft-qsp} package, one needs to enforce the condition $\max_{x \in[-1, 1]}|P(x)|<1$; hence, we rescale the EFP as given in Eq.~\eqref{eq:minimax-filter} by a constant factor of 0.999.

\begin{figure*}
\centering
\includegraphics[width=0.6\columnwidth]{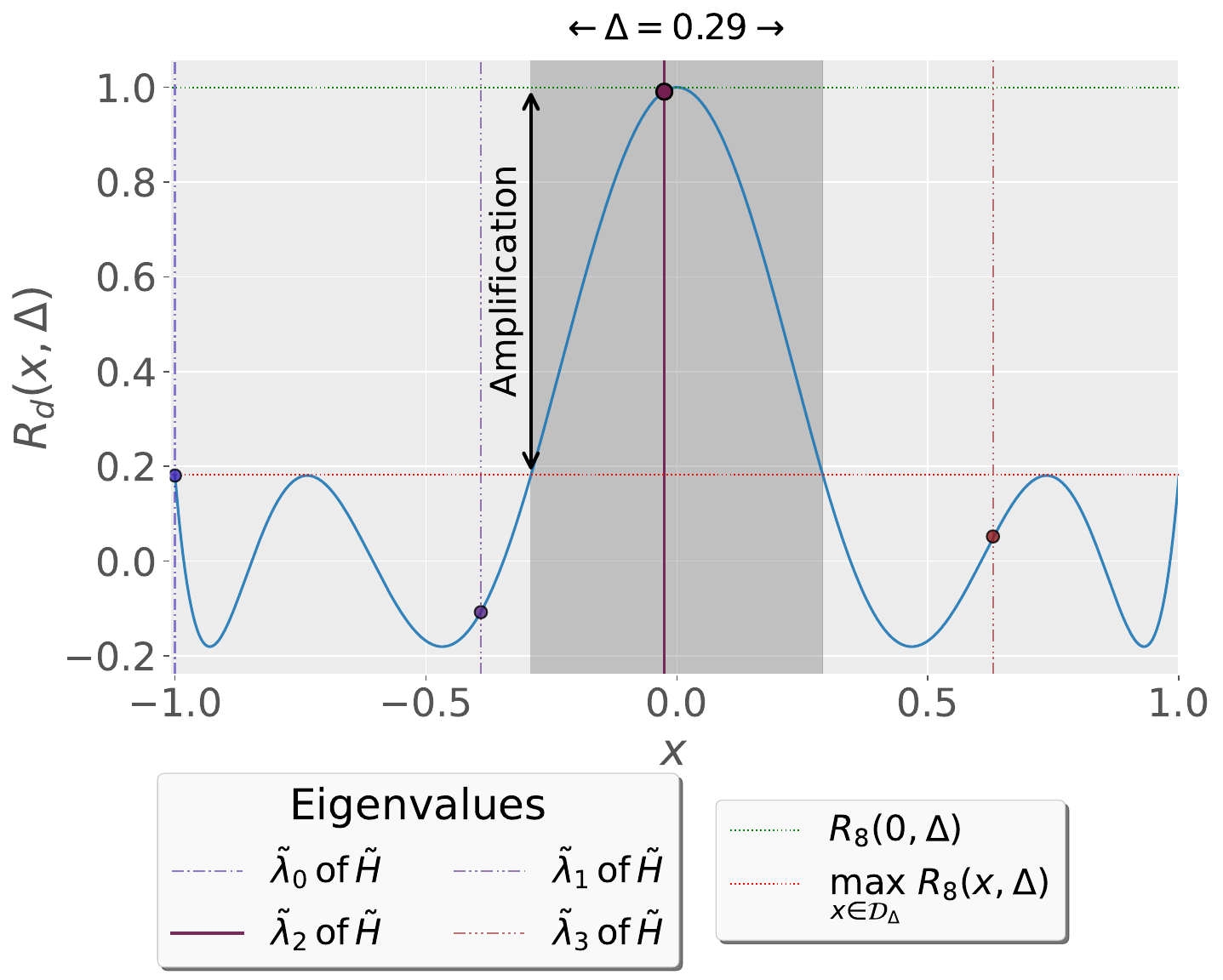}
\caption{Transformation of eigenvalues $\{\tilde{\lambda}_i\}$ of a shifted-and-inverted matrix $\tilde{H}$ by an eigenstate filtering polynomial of degree $d = 8$ through QSVT.
Eigenvalues are transformed as $\{R_8(\tilde{\lambda}_i, \Delta)\}$, as indicated by the intersection points.
The one closest to the origin, $\tilde{\lambda}_2$ is amplified the most.} 
\label{plot:qsvt_demo}
\end{figure*}

As an illustrative example, we consider the same rescaled and shifted \ce{H2} Hamiltonian as in the Cheb-Inv example (Fig.~\ref{fig:chebdecomp}) and show the polynomial transformation of its eigenvalues using the QSVT framework in Fig.~\ref{plot:qsvt_demo}, where we consider the eigenstate filtering polynomial of degree $d = 8$.
The eigenvalue closest to the spectral origin (in this case $\lambda_2$) is amplified the most compared to other eigenvalues, for any given polynomial degree, even low-order ones.
The ratio by which the target eigenstate can be maximally amplified compared to the rest of the spectrum is bounded by: 
\begin{equation}
    \rho_{amp} = \frac{R_d(0, \Delta)}{\max_{x \in \mathcal D_{\Delta}}R_d(x, \Delta)}. 
\label{amp_r}
\end{equation}
This quantity characterizes the intrinsic amplification capability of a degree-$d$ EF polynomial in the limit that the target eigenvalue is mapped to the origin. 
In practice, when $\omega \neq \lambda_t$ and the shifted spectrum is spaced such that there are some eigenstates lying inside the spectral window $\Delta$, the effective amplification is given by 
\begin{equation}
    \tilde{\rho}_{amp} = \frac{R_d(\tilde{\lambda}_t, \Delta)}{\max_{j \neq t}R_d(\tilde{\lambda}_j, \Delta)}, 
\label{amp_r_effective}
\end{equation}
which reduces to Eq.~\eqref{amp_r} when $\tilde{\lambda}_t \rightarrow 0$ and all non-target eigenvalues lie outside $\Delta$. 

\subsection{Eigenstate Filtering-Based Quantum Inverse Power Iteration Algorithm}

\begin{figure*}[t]
\centering
\includegraphics[width= 17cm, height=5cm]{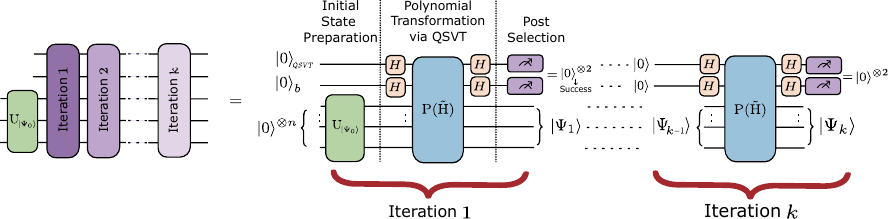}
\caption{Illustration of EFP-based QIPI. After initializing the guess state, each QIPI iteration $k$ involves two main blocks: A polynomial transformation through QSVT, which involves an alternate sequence of $U_{\tilde{H}}$ and $U_{\tilde{H}}^{\dagger}$ with sandwiched projectors as shown in Fig.~\ref{fig:qsvt}, and post selection on the 0 subspace to get the next statevector $\ket{\Psi_k}$.}
\label{fig:QIPI_schematic}
\end{figure*}

We now present our EFP-based QIPI algorithm, which uses the QSVT framework to implement the filtering polynomial $R_d(x, \Delta)$.
Given the Hamiltonian $H$, we aim to find any general eigenstate given a guess $\omega$ to the eigenvalue that is already known.
As described in previous sections, we consider a shifted and scaled Hamiltonian $\tilde{H}$. 
We prepare an initial trial state $\ket{\Psi_0}$ having nonzero overlap with the target state via $U_{\ket{\Psi_0}}$. 
Then, we block encode our rescaled and shifted Hamiltonian $\tilde
{H}$ into a larger unitary $U_{\tilde
{H}}$, as given by Eq.~\eqref{eq: qipi_block_en},
and iteratively apply the polynomial transformation $P(\tilde
{H}) = R_d(\tilde
{H}, \Delta)$ through QSVT for sufficient iterations $k$ as outlined in Algorithm~\ref{alg:QIPI} and illustrated in Fig.~\ref{fig:QIPI_schematic}. 
In the preprocessing step,
$s$ can be set based on the spectral norm $||H|| = \max(\lambda_i)$ such that $s$ is upper bounded as $s < ||H|| + \abs{\omega}$.
For block encodings like LCU, $s$ can be roughly estimated with the 1-norm of Pauli coefficients $\alpha$, such that  $s = \alpha + |\omega|$, without requiring any knowledge of the spectrum \cite{Lin_2020}; however, these assumptions, by definition, overestimate $s$, and the spectrum might not be as widely separated as before, leading to an increase in the polynomial degree required for the convergence.
Alternatively, a tighter estimate than the 1-norm might be obtained using approximate classical calculations; for instance, variational methods can be used to estimate the spectral range.
The convergence of EFP-based QIPI depends on the spectral gap $\Delta$ and the degree $d$ of the polynomial. 
Each filter application amplifies the target state by a factor of $\tilde{\rho}_{amp}$ as given by Eq.~\eqref{amp_r_effective} compared to every off-target component, or $\rho_{amp}$ as given by Eq.~\eqref{amp_r} when all the off-target states are outside $\Delta$.
In that latter case, off-target components are suppressed by at least the theoretical bound $2e^{-d\Delta/\sqrt{2}}$.
Hence the error converges as $O((2e^{-d\Delta/\sqrt{2}})^k)$ in the worst case or as $\rho_{amp}^k$ in the best case.

After each QIPI iteration $\ket{\Psi_k} = P(\tilde
{H})\ket{\Psi_{k-1}}$, we post-select on $\ket{0}_{QSVT}$ and $\ket{0}_b$ ancilla to keep the good branch that has the correct block encoding and discard all the other outcomes. 
The ancilla overhead for EFP-based QIPI is constant due to the use of QSVT, but the success probability is state-dependent. 
For one filter application, the success probability is
$p_k = \lVert R_d(\tilde{H}, \Delta)\ket{\Psi_{k-1}} \rVert^2$
which depends on the spectral distribution of $\ket{\Psi_{k-1}}$. 
Since we post-select after each iteration, the total success probability after $k$ steps is the product of per-step success probabilities.
We give detailed analysis of success probabilities for the systems considered in this work in Sec.~\ref{subsub: succ-prob}.
For proof of principle demonstration, we employed the post selection scheme as described above,  but for large-scale applications of the algorithm the probabilistic overhead of QIPI can be reduced with amplitude amplification (AA) at the cost of additional depth and queries by building a spectral reflection of the projector $P_{\lambda_t}$ (see Eq.~\eqref{eq: P_A}) as  $R_{\lambda_t} = 2P_{\lambda_t} - I$ and wrapping it inside fixed point AA (FPAA)\cite{Lin_2020, yoder2014fixed}.
FPAA can be applied when a lower bound on $\bra{\psi_t}\ket{\Psi_0} = c_t$ is known, and, in contrast to standard AA, it does not suffer from overshooting issues.

We also want to emphasize the fact that the eigenstate filtering polynomial does not approximate the function $(H-\omega I)^{-1}$, but acts as an approximate spectral projector onto the eigenspace closest to the shift. 
But we nevertheless refer to it as a realization of quantum inverse power iteration, because the eigenstate filtering polynomial has the following properties: similar to the inverse function $1/x$, the EFP also peaks at the origin and decays as we move away from the origin.
After $k$ iterations, the amplitude of the target eigenstate relative to the off-target is amplified by the factor $\tilde{\rho}_{amp}^k$.
Thus, although the EFP does not approximate $(H-\omega I)^{-1}$, it reproduces the relative-amplification mechanism responsible for the geometric convergence of inverse power iteration.
This property governs QIPI's convergence. 
We thus use EFP as a more stable replacement for the truncated inverse, while the iteration loop remains unchanged. 

\begin{algorithm}[H]
\caption{Quantum Inverse Power Iteration}
\label{alg:QIPI}
\begin{algorithmic}[1]
\Require Hamiltonian $H$, shift $\omega$, EFP degree $d$, window $\Delta$, trial state $\ket{\Psi_0}$
\Ensure Target state $\ket{\psi_{t}}$
\State \textbf{Preprocessing:}
\State $s \gets \max\bigl(\lvert \lambda_{\min}-\omega\rvert, \lvert \lambda_{\max}-\omega\rvert\bigr)$ \Comment{Can be approximated with a $||H||$ estimate}
\State Shift-and-scale Hamiltonian: $\tilde
{H} \gets \dfrac{H - \omega I}{s}$ \Comment{$\operatorname{spec}(\tilde{H})\subseteq [-1,1]$}
\State Construct a block-encoding $U_{\tilde{H}}$ of $\tilde {H}$ \Comment{For instance as in Eq.~\eqref{eq: qipi_block_en}}
\State Compute $d+1$ QSVT phase factors $\vec{\phi}$ that implement $P(x)= R_d(x,\Delta)$

\Statex

\State \textbf{Iterative polynomial application:}
\State Prepare $\ket{\Psi_0}$
\For{$k$ iterations}
  \State Apply $P(\tilde
{H})$ via QSVT with angles $\vec{\phi}$ according to Eq.~\eqref{eq:qsvt-even}
  \State Measure the two ancilla qubits
  \State Post-select on the ``good'' subspace $\ket{00}$ \Comment{Alternatively: perform FPAA}
\EndFor
\Statex
\State \textbf{Return:} $\ket{\psi_{t}} \gets \ket{\Psi_k}$. 
\end{algorithmic}
\end{algorithm}

\section{Results and Discussion}
\label{sec:results}
In this section, we present numerical simulations of proof-of-principle applications of QIPI to target molecular excited states.
We divide our analysis into two parts: the first demonstrates exact applications of QIPI, and the second benchmarks QIPI by decomposing the arbitrary phase rotations required to perform QSVT into a universal gate set, namely the Clifford + T set, and provides resource estimates to assess its feasibility on FTQC architectures.
We additionally compare the query complexity of EFP-based QIPI with existing methods.
Finally, we examine QIPI using chemically informed, symmetry-adapted initial states and analyze the corresponding per-iteration and cumulative post-selection success probabilities.

\subsection{Simulation Details}
We test the performance of EFP-based QIPI on three small molecular systems where exact diagonalisation as a reference is feasible, namely \ce{H2}, \ce{LiH}, and \ce{BeH2}.
For all three systems, the fermionic Hamiltonian is mapped to qubits with the parity mapping,\cite{seeley2012bravyi} which results in a standard two-qubit reduction, leaving us with 2, 10, and 12 qubits, respectively.
The H-H bond distance for \ce{H2} is set to 0.7414 \AA. 
The Li-H bond distance in \ce{LiH} and the Be-H bond distance in \ce{BeH2} are set to 1.6 \AA $ $ and 1.326 \AA, respectively. 
All calculations are performed using canonical orbitals obtained with \texttt{pyscf}\cite{pyscf} in the STO-3G basis set.\cite{sto-3g}
For demonstration purposes, the spectral shift $\omega$ is chosen such that the desired eigenvalue is the unique eigenvalue nearest to $\omega$. 
Unless stated otherwise, the initial trial state is constructed to satisfy
$\abs{\bra{\psi_t}\ket{\Psi_0}}^2 = 0.2$ to probe convergence with low-overlap initial states as strongly correlated excited states of molecular systems can exhibit Hartree--Fock overlaps significantly below the requirements of many phase-estimation based algorithms.\cite{fomichev2024initial}
We first consider ideal numerical simulations in which the QSVT phase rotations are implemented exactly, and then turn to fault-tolerant estimates obtained by synthesizing each phase rotation into the Clifford+T gate set to a prescribed tolerance $\epsilon$.
To assess the performance of EFP-based QIPI irrespective of the chosen block-encoding strategy as a variety of different schemes could be employed in the algorithm, we treat the block-encoding oracle as exact throughout this section, and its implementation cost is not included in the reported T counts.

\begin{figure*}
\includegraphics[width=\linewidth]{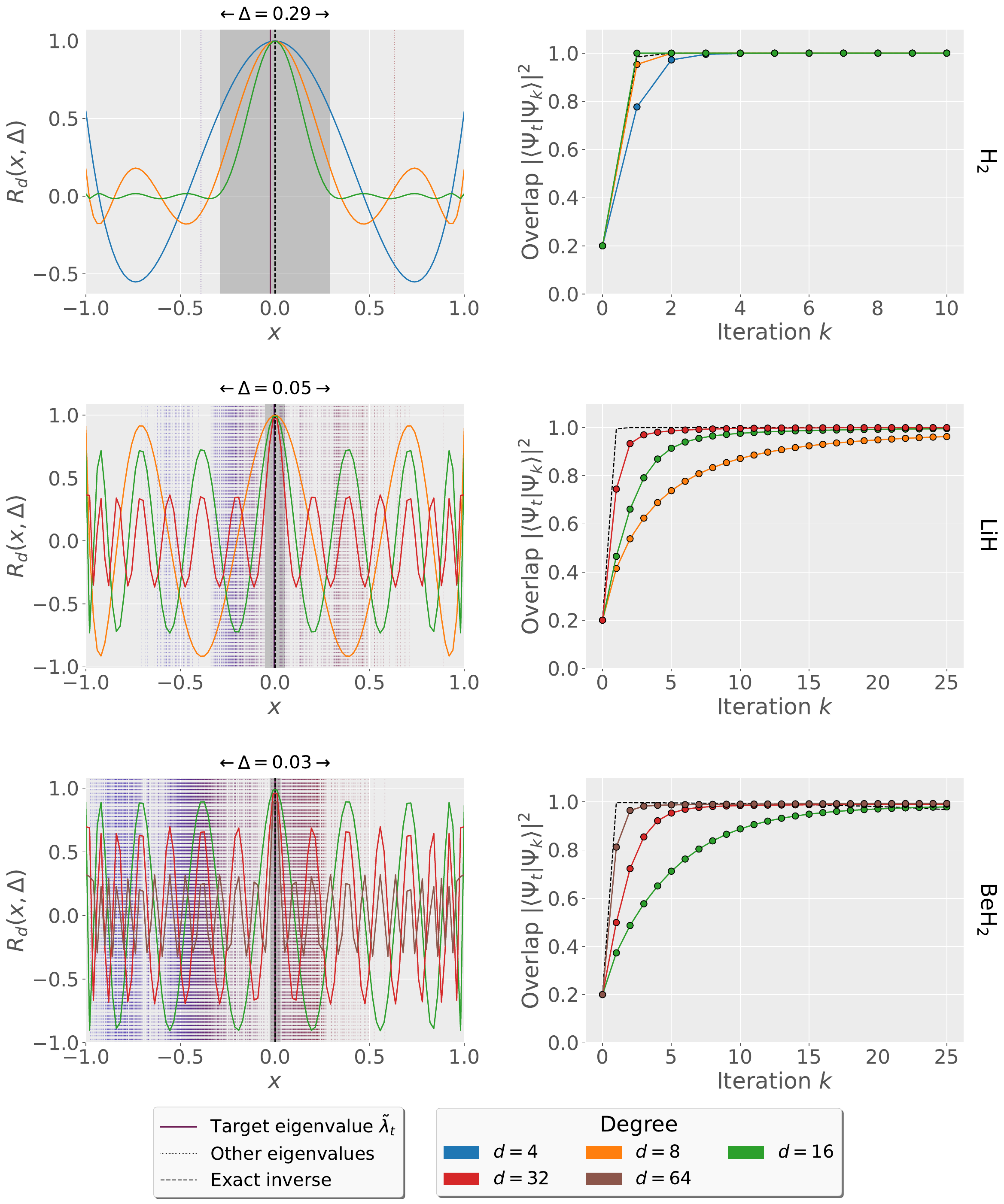}
\caption{QIPI simulation results on \ce{H2}, \ce{LiH}, and \ce{BeH2} with exact QSVT rotations.
(Left) Varying degree EFP transformation of eigenvalues of $\tilde{H}$.
Vertical lines denote the spectrum of $\tilde{H}$, with the target $\tilde{\lambda_t}$ represented in bold.
The eigenvalue closest to the spectral origin is amplified the most for all degrees of the EF polynomial.
(Right) Convergence behaviour of QIPI for the corresponding EF polynomial.
}
\label{fig: QIPI}
\end{figure*}

\subsection{Exact simulations}
\label{sub: exact-simulations}
We first simulate QIPI exactly, i.e., without any circuit decompositions or synthesis overhead of arbitrary rotations.
To compare it with the Cheb-inv approach as presented in Section~\ref{sec: Cheb-Inv IPI}, we apply the EFP-based QIPI to the \ce{H2} Hamiltonian with the same parameters and the appropriate choice of $\Delta$.
As shown in Fig.~\ref{fig: QIPI}, we find that all the eigenstates that lie outside of the spectral gap are efficiently suppressed by EFP-based QIPI irrespective of the poynomial degree.
In comparison to Cheb-in-QIPI, we converge with lower polynomial degrees, and thereby fewer queries to the block encoding of the Hamiltonian, to reach the same accuracy.
Furthermore, we do not observe any divergence issues for the tested instances and shift choices.
As the EF polynomial is symmetric about the origin and maximum target state amplification is achieved for $|\tilde{\lambda}_t - \omega|\rightarrow 0$, EFP-based QIPI is much more robust w.r.t. the energy guess than the decomposition-based approaches, and improved energy estimates accelerate convergence to the target state.

We further test the algorithm on \ce{LiH} and \ce{BeH2} to target arbitrary excited states as shown in Fig.~\ref{fig: QIPI}.
As system size increases, spectral crowding forces one to choose a smaller window $\Delta$, which in turn reduces the achievable target state amplification, and consequently requires higher polynomial degrees and more iterations to suppress unwanted eigenstates.
Nevertheless, QIPI succeeds in converging to the target eigenstate with high fidelity already with modest polynomial degrees for all systems considered here.
Furthermore, in practical applications of the algorithm, properties of the target state can be exploited to reduce the spectral density to that of the target symmetry sector, resulting in larger effective gaps, and thereby smaller required polynomial degrees, as further discussed in Section~\ref{subsec:prac_app}.
For instance, if the aim is to target a specific excited singlet state with QIPI, by preparing a spin-adapted initial state with the appropriate spin and particle number symmetry, the eigenspectrum seen by QIPI is reduced to just the singlet states as can be seen in Fig.~\ref{fig:jw-accesible}.
This can greatly reduce resource requirements as the optimal scaling parameter $s$ and window $\Delta$ will now only depend on the spectral distribution of the singlet states rather than the entire eigenspectrum.
In addition to particle number and spin symmetry, also other properties such as point group symmetries could be exploited to accelerate convergence to the target state through appropriate initial state engineering.

\subsection{Finite-Precision Gate Synthesis Simulations}
We now assess the feasibility and resource requirements of targeting excited states with EFP-based QIPI in a fault-tolerant setting.
Before discussing simulation results, we spend some time on quantifying the cost metric commonly applied in FTQC.  
In gate-based quantum computing, gates are classified in terms of the Clifford hierarchy; one and two-qubit gates such as H, S, and CNOT belong to the Clifford group, while non-Clifford gates such as the T and Toffoli gates lie in higher levels of the hierarchy.
Arbitrary angle rotations around the axes like $R_Z(\phi), R_X(\phi)$, or $R_Y(\phi)$, lie outside this hierarchy altogether and must be synthesized from the gates that form a universal a gate set\cite{nielsen00, cui2017diagonal}, meaning a gate set that can be used to approximate any unitary matrix as a quantum circuit within a given error budget.
The Clifford group together with non-Clifford gates such as the T gate enable such universal quantum computation.\cite{universalcomputation}
In FTQC based on topological codes,\cite{fowler2012surface} Clifford gates can be implemented transversely or absorbed into measurements.\cite{litinski2019game, sunami2025transversalsurfacecodegamepowered}
Implementing T gates needs preparation of high-fidelity T states, which in turn requires expensive protocols such as magic state distillation and cultivation, and they come with substantial qubit- as well as time-overhead.\cite{Litinski_2019, gidney2024magicstatecultivationgrowing}
For this reason, the number of T gates is commonly used as a primary metric to quantify the cost of a fault-tolerant computation.

\begin{figure*}
\centering
\includegraphics[width=0.6\columnwidth]{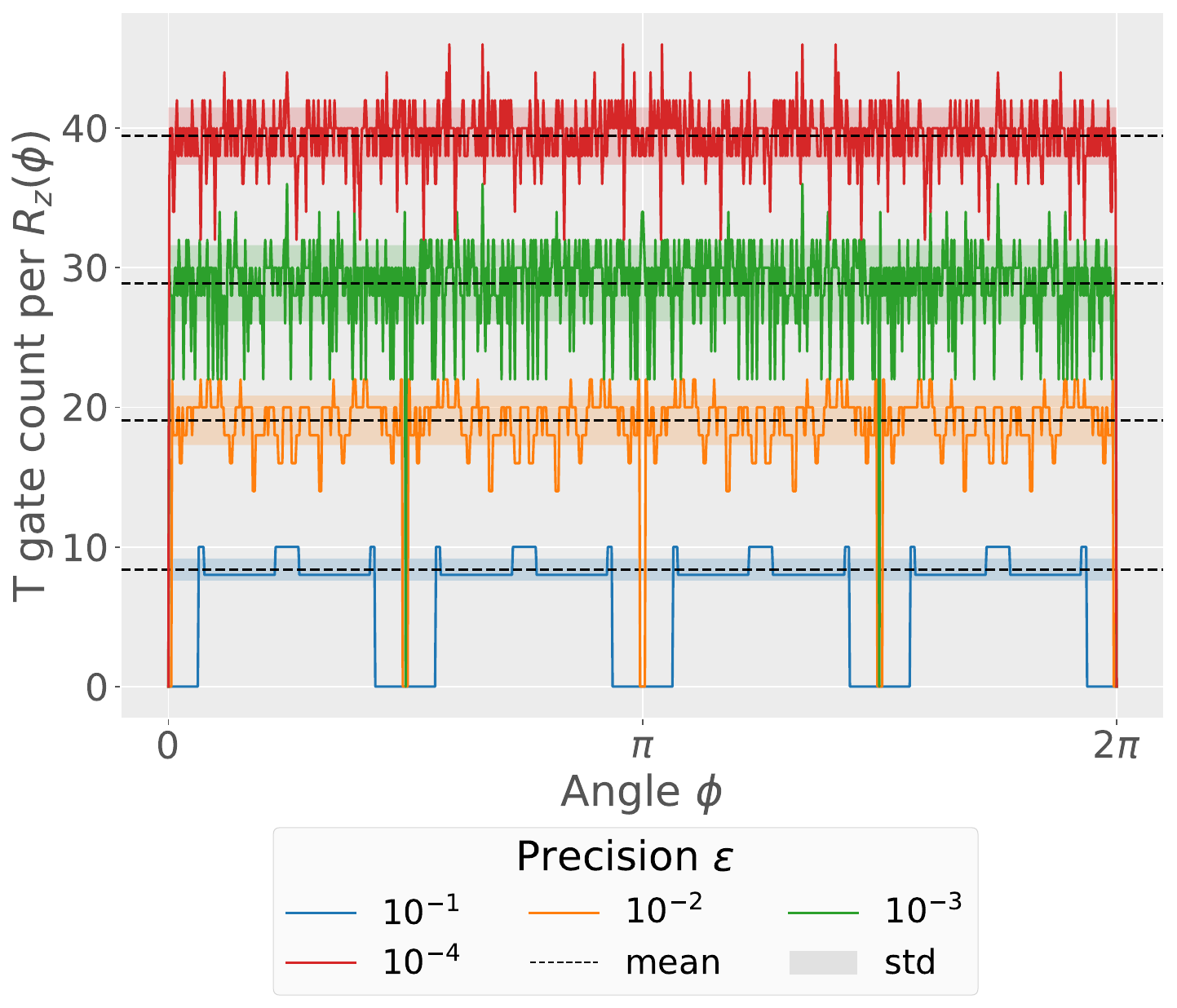}
\caption{Quantitative overview of T gate count per rotation decomposition for different precisions of the arbitrary phase angles $\phi$ of the implemented $R_Z(\phi)$.
Dashed black lines and shaded areas represent mean T gate cost per nontrivial rotation and the standard deviation, respectively.}
\label{plot: t count sweep}
\end{figure*}

QSVT-based algorithms rely on phase-modulated reflections, which are implemented through the application of $d+1$ rotation operations $R_Z(\phi)$ to encode a polynomial of degree $d$.
As mentioned before, such arbitrary rotations can, however, not be implemented directly in FTQC, but they can only be approximated by decomposition into a sequence of Clifford and T gates.
Fig.~\ref{plot: t count sweep} shows the number of T gates required for the ancilla-free synthesis of a $Z$-rotation with arbitrary phase angles decomposed to a given precision $\epsilon$ using \texttt{pygridsynth}.\cite{ross2016optimalancillafreecliffordtapproximation} 
This decomposition strategy achieves a T-count scaling of about $3\log_2(1/\epsilon) + O(\log(\log(1/\epsilon)))$, with each order-of-magnitude improvement in $\epsilon$ costing about an additional 10 T gates for non-trivial phase angles, as evident from Fig.~\ref{plot: t count sweep}.
Thus, there is a trade-off to be made between the accuracy of the rotation-decomposed QSVT implementation and the resulting T-gate count.

\begin{figure*}
\includegraphics[width=\linewidth]{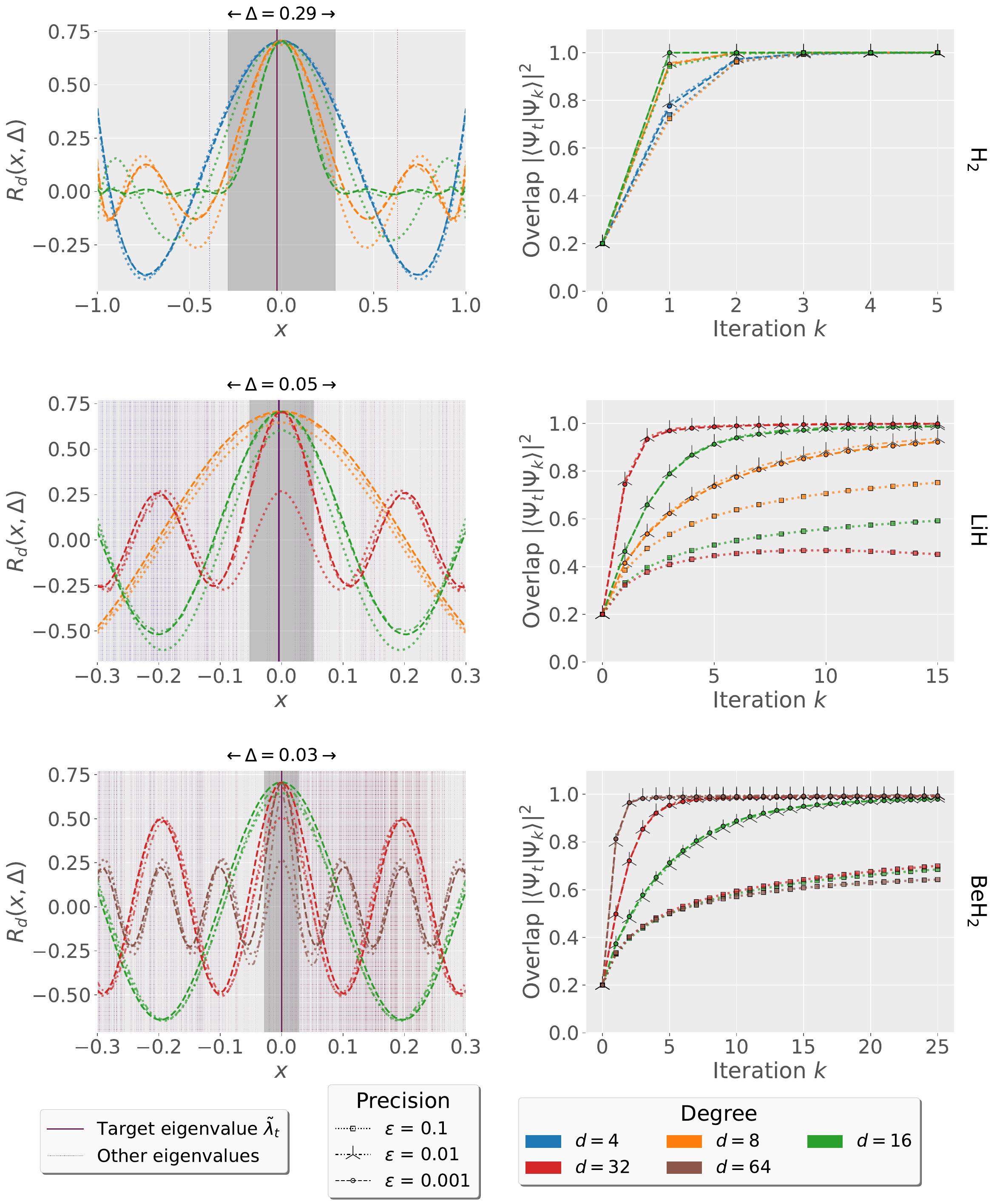}
\caption{QIPI simulation results on \ce{H2}, \ce{LiH}, and \ce{BeH2} with rotation decomposition for different precisions.
(left) EFP transformations with finite-precision rotation synthesis.
For lower precisions, amplification is not retrieved fully; as we go to higher precisions, the amplification effect is preserved.
(right) Corresponding convergence behaviour of QIPI.
}
\label{fig: QIPI_rot}
\end{figure*}

The effect of the approximate EFP application to different molecular Hamiltonians is shown in Fig.~\ref{fig: QIPI_rot} for different rotation precision values.
Low precision prevents the quantum circuit from accurately reproducing the amplification achieved by the EFP, resulting in slower convergence to the target state compared to the exact QIPI application shown in Fig.~\ref{fig: QIPI}.
As the decomposition precision increases, the EF amplification is more faithfully realized, resulting in faster convergence.
Typically, many FT quantum algorithms require high-precision synthesis of single-qubit rotations on the order of $10^{-10}$ ($\approx$ 100 T gates per rotation) to achieve the desired algorithmic accuracy.\cite{qrechem2024}
In our case, however, we find that lower-precision decompositions up to $10^{-3}$ ($\approx$ 30 T gates per rotation) are sufficient for the resulting circuit to preserve the required target state amplification, making QIPI more accessible for the mid-term FTQC regime.

\begin{figure*}[!h]
\centering
\includegraphics[width=0.6\columnwidth]{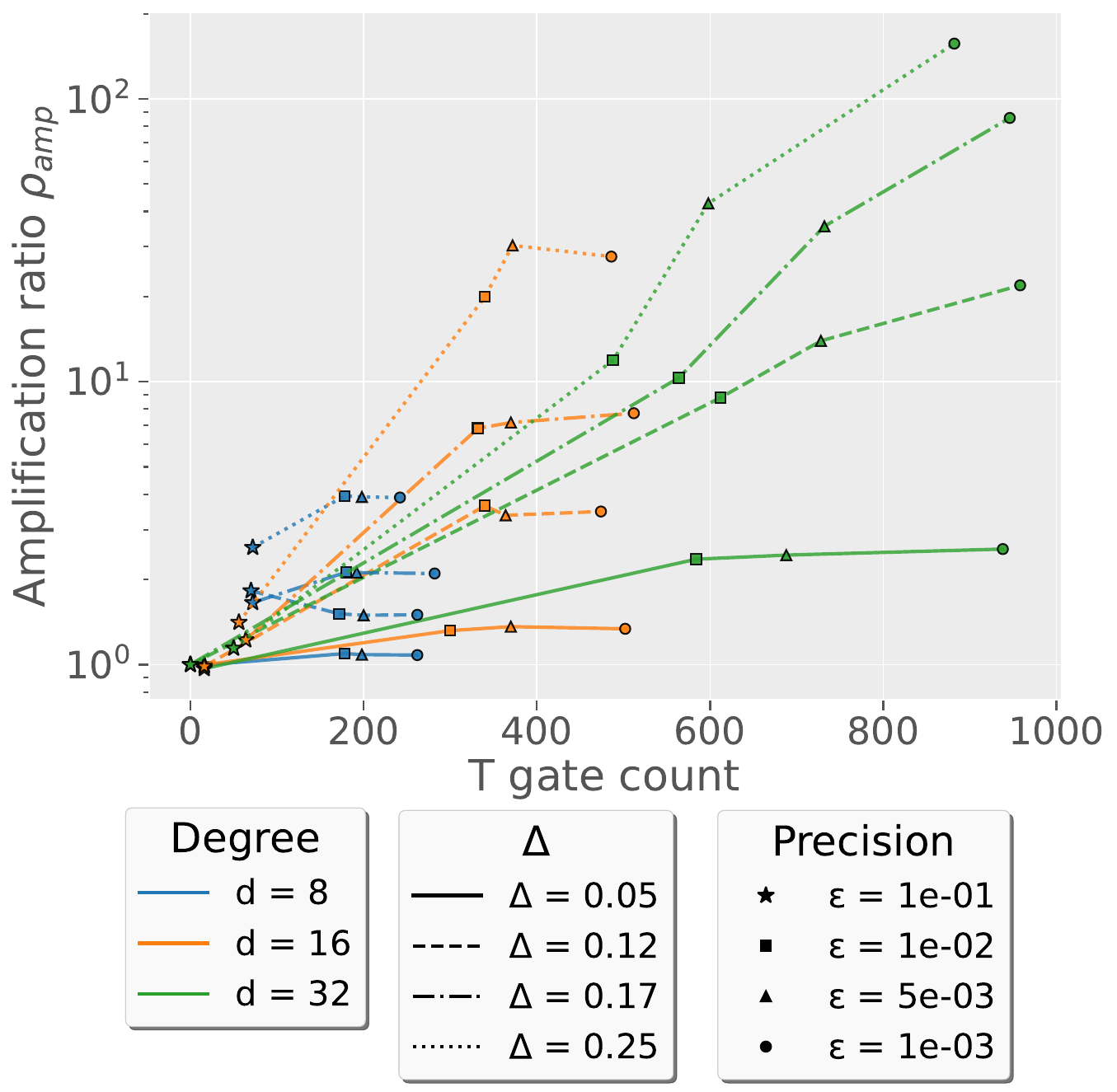}
\caption{Amplification ratio $\rho_{amp}$ of Eq.~\ref{amp_r} versus T gate count for various rotation decomposition precisions $\epsilon$ across different window values $\Delta$ and varying degrees $d$. Larger polynomial degrees only result in greater amplification for high-precision decompositions, thereby increasing the T-gate cost significantly compared to similar amplification with lower-degree lower-precision EFP implementations.
}
\label{plot:tgate}
\end{figure*}

We now analyze the per-iteration resource scaling of EFP-based QIPI in terms of the T-gate count as a function of the polynomial degree $d$, the choice of spectral window $\Delta$, and the precision $\epsilon$ of the single-qubit rotation decomposition. 
As discussed above, convergence of EFP-based QIPI depends on how well a polynomial suppresses off-target components per iteration, and we use the amplification ratio as defined in Eq.~\eqref{amp_r} to quantify it.
Ideally, in a gate-synthesis-free picture, amplification $\rho_{amp}$ improves monotonically with increasing degrees and increasing $\Delta$, but when chemical Hamiltonians having closely spaced spectra are considered, if $\Delta$ is chosen arbitrarily large, off-target eigenvalues can no longer be filtered and amplification instead follows Eq.~\eqref{amp_r_effective}.
Hence an appropriate $\Delta$ value must be chosen based on the estimates of the spectral gaps around the target state, and on properties of the target state itself.
For example, a small system like \ce{H2} naturally admits a larger $\Delta$ owing to its widely spaced spectrum. 
The spectral density can also be reduced by exploiting symmetries, as discussed in Sec.~\ref{sub: exact-simulations}, which would permit a larger effective $\Delta$; for molecular Hamiltonians considered in this work, however, we consider the full spectrum for demonstration purposes. 
Consequently, for the denser spectra of 
\ce{LiH} and \ce{BeH2}, $\Delta$ is typically much smaller than the theoretical upper bound of $1/\sqrt{12}$.
Smaller $\Delta$ values give relatively less suppression of off-target states for a given polynomial degree and hence require higher degree polynomials to reach a substantial amplification, which could make the iterative application of the EFP polynomial sub-optimal as a single application of a larger-degree polynomial might seem more resource-efficient.
However, as soon as we take gate synthesis into account, this picture changes.
A degree-$d$ EFP requires $\mathcal{O}(d)$ single-qubit rotations, each of which must be synthesized into Clifford$+$T set at cost $\sim 3\log_2(1/\epsilon)$ T gates per rotation; the per-iteration T count therefore grows with both $d$ and $1/\epsilon$.
As shown in Fig.~\ref{plot:tgate}, choosing large $\Delta$ does improve $\rho_{amp}$, but such a choice is not admissible for the crowded spectra
of the molecular Hamiltonians considered here, such as \ce{LiH} and \ce{BeH2}.
The regime of very small $\Delta$ with varying degrees $d$ yields only marginal amplification, as expected. 
As seen in Fig.~\ref{fig: QIPI_rot}, the amplification factor of the polynomial is fairly recovered at $\epsilon = 10^{-3}$, 
which fixes the $\epsilon$ budget, and the only way to achieve the substantial suppression of off-target eigenstates is to increase the degreee of the polynomial, which in turn increases the number of T gates required. 
Such high-degree coherent circuits also increase the number of block encodings required per run, which in turn demand multiple layers of T state distillation to reach required algorithmic accuracy, further adding to the overall fault-tolerant implementation cost.\cite{litinski2019game}
The iterative nature of QIPI is beneficial in such a scenario as the overall implementation cost is set by the space-time volume of the circuit. Iterative application of the EFP minimizes this space-time volume and hence requires smaller distance error correcting codes/smaller distillation overhead\cite{fowler2012surface, litinski2019game}.
This cost grows even higher when early/mid fault-tolerant regimes are considered, where there is a presence of significant residual noise, and applying such deep coherent circuits becomes practically infeasible.
Therefore, using very high degree polynomials does not necessarily provide optimal suppression of off-target elements when associated synthesis overhead is taken into account.

Since smaller $\Delta$ values come with a high degree, subsequently higher T gates and associated fault-tolerant resource overhead, this trade-off is precisely what motivates applying a low-order EFP iteratively rather than a single high-order one.
In QIPI, the per-iteration amplifications multiply, therefore, the same overall amplification can be reached with many shallow, low-T circuits instead of one deep, high-degree circuit, which also brings a further benefit when noisy scenarios are considered.
In the early/mid fault-tolerant regime, using one coherent high-degree circuit of large depth, noise accumulation can scramble the information entirely, but with QIPI when a state at the $k$th iteration drifts away due to insufficiently suppressed
hardware errors, those errors can be partially mitigated through the iterative nature of the algorithm as long as the overlap of the drifted state does not decrease significantly. 
Hence, a low-order iterative scheme such as in QIPI is practically more robust than a single high-order polynomial application.
However, it might not be optimal to choose the lowest possible degree to optimize the implementation cost, as this heavily depends on the choice of window $\Delta$. 
The theoretical upper bound by which off-target states can be suppressed, i.e., $2 e^{-\sqrt{2}\ell \Delta}$, only predicts suppression when $d\Delta> \sqrt{2} \ln 2$.
Thus, for a fixed $\Delta$, only polynomial degrees above this threshold are guaranteed to suppress the off-target components under this theoretical bound.
For instance, the \ce{LiH} example with $\Delta = 0.05$ shown in Figs.~\ref{fig: QIPI} and \ref{fig: QIPI_rot} in theory requires at least degree $d > 19$ to efficiently suppress off-target elements, while in practice convergence is achieved already with lower degrees due to the analytical bound being loose.

\subsection{Query Complexity Analysis}
Here we derive query complexity of EFP based QIPI and compare it against existing eigenstate filtering methods, followed by a numerical analysis of query complexity with different initial state overlaps.
Any initial state $\ket{\Psi_0}$ can be expanded in the eigenbasis of the Hamiltonian as $\ket{\Psi_0} = \sum_i c_i \ket{\psi_i}$ with coefficients $c_i = \langle \psi_i | \Psi_0 \rangle$.
We assume a non-zero overlap with the target eigenstate $\ket{\psi_t}$, and denote its magnitude by $\gamma = |c_t| = |\langle \psi_t | \Psi_0 \rangle|$.
Absorbing a global phase so that the target coefficient is real and positive, we rewrite this as
\begin{equation}
\ket{\Psi_0} = c_t\ket{\psi_t} + \sum_{i \neq t} c_i \ket{\psi_i}.
\end{equation}

Applying a filtering polynomial $P_d(\tilde{H})$ that satisfies the spectral filtering properties (e.g.\ the EFP given in Eq.~\eqref{eq:minimax-filter}) gives
\begin{equation}
\begin{aligned}
\ket{\Psi_1} = P_d(\tilde{H})\ket{\Psi_0} 
&= c_t P_d(\tilde{\lambda}_t)\ket{\psi_t}
+ \sum_{i \neq t} c_i P_d(\tilde{\lambda}_i)\ket{\psi_i} \\
&= P_d(\tilde{\lambda}_t)\left(
c_t \ket{\psi_t}
+ \sum_{i \neq t} c_i \frac{P_d(\tilde{\lambda}_i)}{P_d(\tilde{\lambda}_t)} \ket{\psi_i}
\right).
\end{aligned}
\end{equation}

Thus, relative to the target component, each off-target component is compressed by a factor 
$\left|\frac{P_d(\tilde{\lambda}_i)}{P_d(\tilde{\lambda}_t)}\right| \le \tilde{\rho}_{comp}$, where 
$\tilde{\rho}_{comp} = \frac{1}{\tilde{\rho}_{amp}} = \frac{\max_{i \neq t} |P_d(\tilde{\lambda}_i)|}{|P_d(\tilde{\lambda}_t)|}$ from Eq.~\eqref{amp_r_effective}.

If $\tilde{\lambda}_j$ is the next closest eigenvalue to $\omega$, $\tilde{\rho}_{comp} = \frac{ |P_d(\tilde{\lambda}_j)|}{|P_d(\tilde{\lambda}_t)|}$, and after $k$ applications,
\begin{equation}
\ket{\Psi_k} = P_d(\tilde{H})^k\ket{\Psi_0}
= c_t P_d(\tilde{\lambda}_t)^k \left(
\ket{\psi_t} + \frac{c_j}{c_t}\tilde{\rho}_{comp}^k \ket{\psi_j}
+ \sum_{i \neq t, j} \frac{c_i}{c_t} \left(\frac{P_d(\tilde{\lambda}_i)}{P_d(\tilde{\lambda}_t)}\right)^k \ket{\psi_i}
\right).
\end{equation}
Hence, the relative weight of each off-target component is bounded by $\tilde{\rho}_{comp}^k/|c_t| = \tilde{\rho}_{comp}^k/\gamma$.
To achieve accuracy $\eta$, we require $\tilde{\rho}_{comp}^k/\gamma \le \eta$, implying 
\begin{equation}
   k =
O\!\left(\frac{\log\left(1/(\gamma\eta)\right)}{\log(1/\tilde{\rho}_{comp})}\right) 
\label{eq:k_scaling}
\end{equation}
iterations to converge; for a fixed non-zero overlap $\gamma$, this reduces to
$k = O\!\left(\frac{\log(1/\eta)}{\log(1/\tilde{\rho}_{comp})}\right)$.

QSVT provides an efficient framework to implement $P_d(\tilde{H})$, as the degree-$d$ EF polynomial can be applied using $O(d)$ queries to the block-encoding of $\tilde{H}$.
Therefore, the total query complexity is
$Q_{\mathrm{total}} = O(dk)$.
For the EFP, the compression factor satisfies $\tilde{\rho}_{comp} \le \exp(-\Theta(d\Delta))$, where $\Delta$ is the spectral gap around the target eigenvalue.
Consequently, $\log(1/\tilde{\rho}_{comp}) = \Theta(d\Delta)$, and
$Q_{\mathrm{total}} = O\!\left(d \cdot \frac{\log(1/\eta)}{d\Delta}\right) = O\!\left(\frac{1}{\Delta}\log\frac{1}{\eta}\right)$,
up to constant factors. 
The overall complexity is thus governed by the spectral gap and target accuracy.
The same reasoning applies to the proposed EFP-based QIPI algorithm.
Although QIPI performs repeated applications of a fixed-degree polynomial, each step incurs $O(d)$ query cost and achieves the same compression factor $\tilde{\rho}_{comp}$ at the end.
As a result, the total query complexity remains $O\left(\frac{1}{\Delta}\log\frac{1}{\eta}\right)$, conditional on successful postselection.

\begin{table}[ht]
\centering
\begin{tabular}{p{7 cm}|p{3.4 cm}|p{1.5 cm}|p{3 cm}}
\hline
\hline
Algorithm & Queries to block encoding & Queries to $U_{\ket{\Psi_0}}$ & No. of ancilla qubits \\
\hline
Eigenstate filtering (QPE + AA)\cite{QPEQuery} & $O(\tfrac{1}{\gamma^2 \Delta \eta})$ & $O(1/\gamma)$ & $O(\log (1/(\eta \Delta)))$ \\
\hline
Poulin and Wocjan \cite{FilteringPoulin} & $O(\tfrac{1}{\gamma \Delta \eta})$ & $O(1/\gamma)$ & $O(\log (1/(\eta \Delta)))$ \\
\hline
LCU-based filtering \cite{Childs_2017} & $O(\frac{1}{\gamma \Delta}\log\frac{1}{\eta})$ & $O(1/\gamma)$ & $O(\log\log (1/\eta) + \log(1/\Delta))$ \\
\hline
QFSM + AA \cite{khinevich2025quantumpoweriteration} & $O(\tfrac{1}{\gamma (\Delta^2 - \omega^2)}\log\frac{1}{\gamma^2\eta})$ & $O(1/\gamma)$ & $O(\log\log (1/\eta) + \log(1/\Delta))$ \\
\hline
Polynomial filtering + AA\cite{Lin_2020} & $O(\tfrac{1}{\gamma \Delta}\log\frac{1}{\eta})$ & $O(1/\gamma)$ & $3$ \\
\hline
QETU (only for ground state)\cite{dong2022groundstate} & $O(\tfrac{1}{\gamma^2 \Delta}\log\frac{1}{\eta})$ & $O(1/\gamma^2)$ & $1$ \\
\hline
\textbf{EFP-based QIPI} (this work) & $O\!\left(\tfrac{1}{\gamma^2\Delta} \log\frac{1}{\eta}\right)$ & $O(1/\gamma^2)$ & 2 \\
\hline
\textbf{EFP-based QIPI + AA} (this work) & $O\!\left(\tfrac{1}{\gamma\Delta} \log\frac{1}{\eta}\right)$ & $O(1/\gamma)$ & 3 \\
\hline
\end{tabular}
\caption{Comparison of query complexities of various eigenstate filtering algorithms. Here $\Delta$ is the spectral gap, $\gamma$ the magnitude of the initial overlap with the target state, $\eta$ the required accuracy, and $\omega$ the energy guess.}
\label{table}
\end{table}

Amplitude amplification can be used to boost the success probability, leading to an additional factor of $O(1/\gamma)$ in the query complexity, without amplitude amplification, this factor worsens to $O(1/\gamma^2)$.\cite{Lin_2020}
This matches with the near-optimal complexity of the QSP-based eigenstate filtering algorithm, which uses the same EFP construction but applies it only once rather than iteratively.\cite{Lin_2020}
With amplitude amplification, this scheme achieves a query complexity of $O\left(\tfrac{1}{\gamma\Delta}\log\frac{1}{\eta}\right)$ and a
constant ancilla overhead of $3$ qubits due to use of QSP and AA.
As discussed in the previous section, the iterative nature of QIPI make it suitable for FTQC applications accounting for finite-precision gate synthesis and associated T gate overhead.
Alternative filtering methods previously proposed in literature result in similar query complexities, but often require a more significant ancilla count.
A widely used literature method is QPE combined with AA, which achieves query complexity $O(\tfrac{1}{\gamma^2 \Delta \eta})$ with an ancilla scaling of $O(\log (1/(\eta \Delta)))$.\cite{QPEQuery}
The filtering method of Poulin and Wocjan achieves a scaling of $O(\tfrac{1}{\gamma \Delta \eta})$ with the same ancilla overhead.\cite{FilteringPoulin}
LCU approaches based on Fourier or Chebyshev expansions implement polynomial filters with query complexity $O\!\left(\frac{1}{\gamma\,\Delta}\log\frac{1}{\eta}\right)$ at an ancilla cost of $O(\log\log (1/\eta) + \log(1/\Delta))$.\cite{Childs_2017}
An alternative low ancilla overhead QSVT-inspired filtering approach, named QETU, attains near-optimal ground-state preparation with a single extra ancilla using controlled time-evolution as the input unitary rather than a block encoding.\cite{dong2022groundstate}
All of the above methods are primarily developed for ground-state preparation, but they could in principle be extended to target low-lying excited states through spectral shifting or windowing techniques.
A more closely related approach, conceptually similar to our QIPI is the quantum folded-spectrum method (QFSM), which also targets excited states directly through generalized quantum signal processing. 
With amplitude amplification, QFSM achieves a query complexity of $O\left(\tfrac{1}{\gamma(\Delta^2-\omega^2)}\log\frac{1}{\gamma^2\eta}\right)$ and an ancilla overhead of $O(\log\log(1/\eta)+\log(1/\Delta))$, somewhat worse than that of EFP-based QIPI.\cite{khinevich2025quantumpoweriteration}
We summarize the query complexities of different filtering methods compared to our QIPI approach in Table \ref{table}.

We see that, compared to other methods, EFP-based QIPI matches the near-optimal query complexity while targeting arbitrary excited states.
More importantly, it does so with comparatively shallow, low-degree iterative circuits and a constant ancilla overhead, rather than the deep coherent circuits the other methods require. 
Note that QIPI is a fault-tolerant algorithm; its iterative nature, however, also makes it well suited for early/mid fault-tolerant regimes, where algorithms try to minimize the depth of long, coherent circuits at the cost of many shorter circuits and extra measurement overhead.\cite{lin2022heisenberg, ding2023even, dong2022groundstate}

\begin{figure*}[t]
\centering
\includegraphics[width=\linewidth]{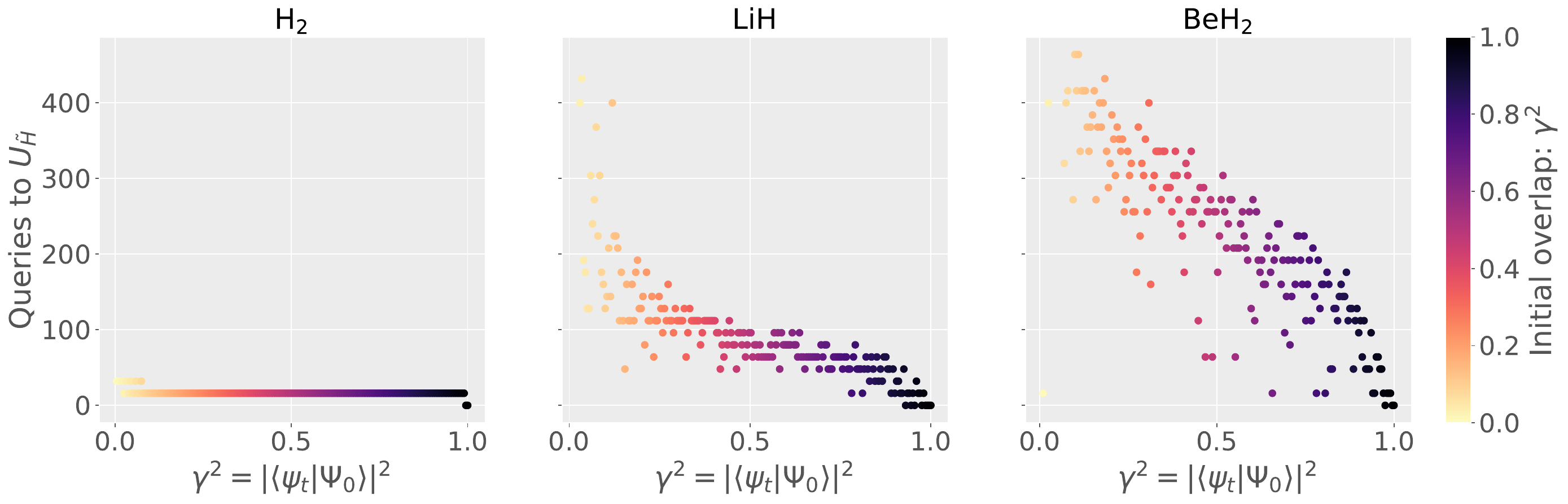}
\caption{Number of queries to reach chemical accuracy ($\eta_c = 1.59$~mHartree) as a function of the squared initial-state overlap $\gamma^2$, using an EFP of degree $d = 16$.}
\label{fig:que}
\end{figure*}
Having established the expected query scaling of QIPI as $Q_\text{total}=O(dk)$ for $k$ iterations, we now verify the query scalings numerically for the molecular systems considered here with EFP-based QIPI.
Fig.~\ref{fig:que} shows the number of queries required by QIPI to reach chemical accuracy of $\eta_c = 1.59$ mHartree
as a function of the squared initial state overlap $\gamma^2$.
We report simulation results for a $d = 16$ degree EFP polynomial.
For \ce{H2}, QIPI reaches chemical accuracy in a single iteration (16 queries) for all but the smallest initial overlaps, consistent with the convergence behaviour seen in Fig.~\ref{fig: QIPI}; since all off-target eigenvalues of \ce{H2} lie well outside the chosen $\Delta$, $\tilde{\rho}_{comp}$ is close to zero, producing the flat curve in Fig.~\ref{fig:que}.
For \ce{LiH} and \ce{BeH2}, the spectra are much denser and the effective compression per iteration is weaker, thus requiring more iterations: the query count grows as the initial overlap decreases, reaching up to $\sim 430$ and $\sim 460$ queries, respectively, at small overlaps.
In all three cases the measured query counts grow only logarithmically with $1/\gamma$, in agreement with the iteration scaling of Eq.~\eqref{eq:k_scaling}, with a prefactor set by $1/\log(1/\tilde{\rho}_{comp})$; for \ce{BeH2} this prefactor is large enough that the logarithmic curve appears nearly linear over the plotted overlap range.
Note that the query counts in Fig.~\ref{fig:que} are conditional on successful post-selection, i.e., they do not include the repetition overhead of $O(1/\gamma)$ (with amplitude amplification) or $O(1/\gamma^2)$ (without) discussed above.

\subsection{Practical applications of QIPI}
\label{subsec:prac_app}
So far, we have analysed QIPI convergence under both ideal polynomial application and finite-precision rotation synthesis, as shown in Figs.~\ref{fig: QIPI} and~\ref{fig: QIPI_rot}, respectively, using randomly generated initial states.
While such states enable a controlled analysis of convergence, they may not represent practically motivated state-preparation scenarios.
Here, we investigate how chemically informed initialization and spectral rescaling affect QIPI convergence.
In particular, we consider initial states with fixed particle number and spin symmetry, and compare them with
unstructured random states and idealized energy-localized states. 
We also examine how restricting the relevant spectrum to the symmetry sector accessible from the initial state permits a tighter scale factor $s$ and, consequently, a larger normalized filtering window $\Delta$.
For the latter, we found the theoretical limit $1/\sqrt{12}$ divided by the scaling parameter $s$ to be a suitable setting to optimally suppress off-target eigenstates in practice, as it maximizes $\Delta$ while taking the spectral spread into account through the spectral scaling parameter.

We now consider three different initial states, namely spin-adapted states, energetically concentrated states, and random superpositions as before.
We construct spin-adapted initial states as configuration state functions (CSFs) of $N$ electrons in $n$ spatial orbitals with a given total spin and spin projection quantum number, and denote them by $\ket{\Psi_{\mathrm{SA}}}$.
As a model of an energetically pre-filtered (``cooled'') initial state $\Psi_{\mathrm{cool}}$ we apply an ideal Gaussian energy filter of width $\sigma$, centered on the shift $\omega$ to the eigenstates $|\phi_i\rangle$ of $H$, with energies $\lambda_i$ each:
\begin{equation}
  |\Psi_{\mathrm{cool}}\rangle = \frac{1}{\mathcal{N}} \sum_i
  e^{-(\lambda_i-\omega)^2/(4\sigma^2)} \, |\phi_i\rangle,
  \qquad
  \mathcal{N}^2 = \sum_i e^{-(\lambda_i-\omega)^2/(2\sigma^2)},
\end{equation}
so that the weights $|\langle\phi_i|\Psi_{\mathrm{cool}}\rangle|^2 \propto
e^{-(\lambda_i-\omega)^2/(2\sigma^2)}$ follow a Gaussian energy distribution of width $\sigma$.
Since every eigenstate enters the superposition, $|\Psi_{\mathrm{cool}}\rangle$ is
not symmetry adapted: it also contains eigenstates of other spin and particle-number sectors which lie close in energy to the target state.
To ensure a fair comparison, we construct the initial states such that the ratio $R_{\mathrm{init}}$ between their squared overlap with the target eigenstate ($c_t$) and their largest squared overlap with any off-target eigenstate ($\max(c_i, i \neq t)$) is held constant:
\begin{equation}
  R_{\mathrm{init}}
  = \frac{c_t}{\max(c_i, i \neq t)}
  = \frac{|\langle\psi_t|\Psi_0\rangle|^2}
  {\displaystyle\max_{i\neq t}|\langle\psi_i|\Psi_0\rangle|^2}.
\end{equation}
Due to this construction, different initial states may have different starting initial overlap but it ensures that all three initial states have the same initial spectral selectivity toward the target.

\begin{figure*}[t]
\centering
\includegraphics[width=\linewidth]{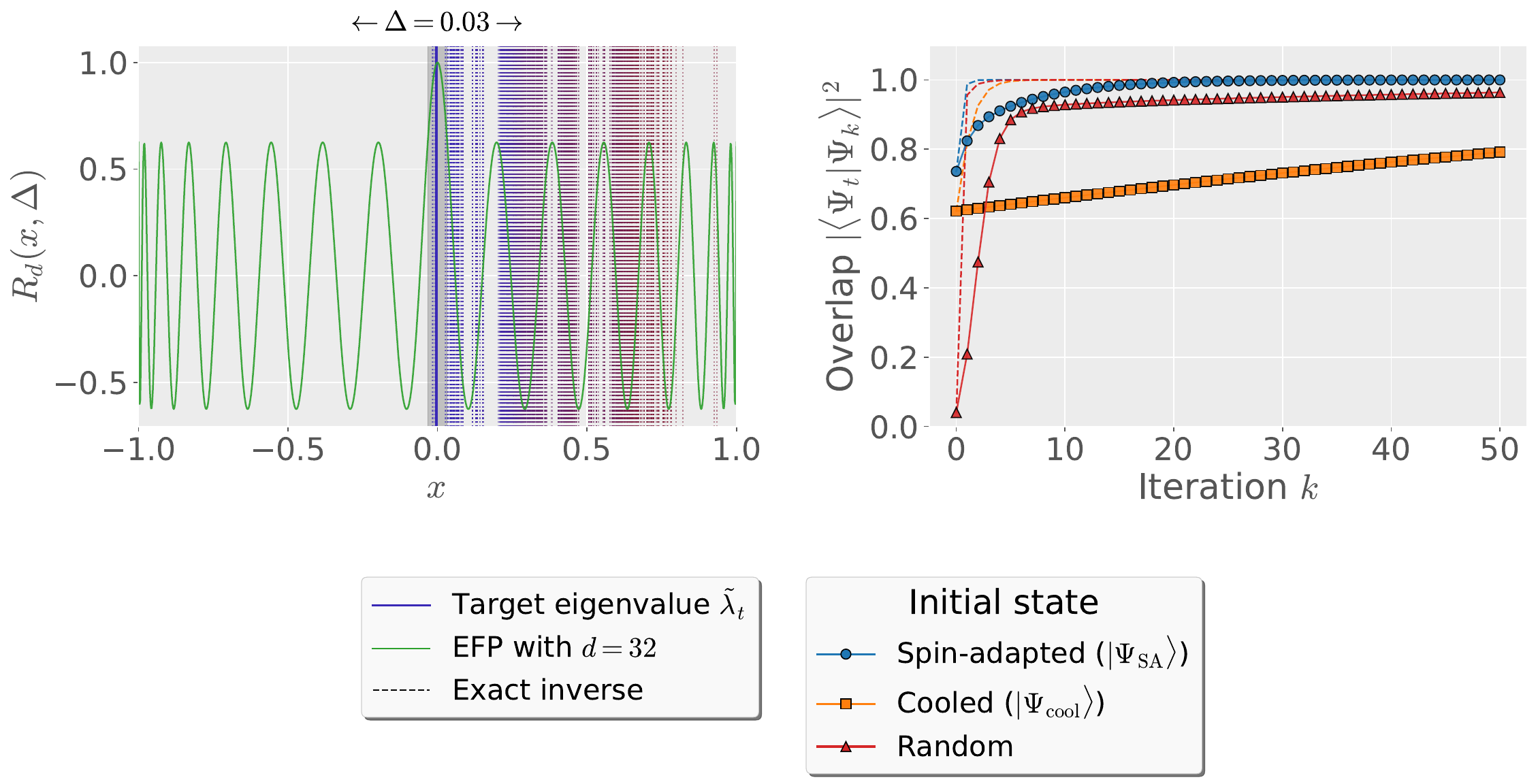}
\caption{Effect of initial-state structure on EFP-based QIPI for
parity-mapped \ce{LiH} in the STO-3G basis.
(left) Degree-$d=32$ eigenstate-filtering polynomial applied to the shifted and scaled Hamiltonian. Vertical lines denote the eigenvalues of $\tilde{H}$, with the target eigenvalue $\tilde{\lambda}_t$ shown as a solid line; the shaded region indicates the filtering window $\Delta=0.03$.
(right) Target-state overlap as a function of QIPI iteration for a spin-adapted triplet CSF, an idealized Gaussian energy-localized (cooled) state, and a random state. 
The three states have the same $R_{\mathrm{init}}$ but initial
target overlaps of $0.74$, $0.62$, and $0.04$, respectively. 
Dashed curves show the corresponding exact inverse iterations. 
}
\label{fig:initial-state-comparison}
\end{figure*}

We again consider the parity-mapped Hamiltonian of \ce{LiH}, but with the \ce{Li}-\ce{H} bond distance set to 3 \AA. 
We choose this setting to reflect more realistic application of our QIPI algorithm to a correlated system with more multiconfigurational character, and consequently lower mean-field state overlap.
We compare the QIPI convergence for the
different initial states in Fig.~\ref{fig:initial-state-comparison}. 
The scaling parameter is chosen as the one-norm, $s = \alpha$, the shift $\omega$ as the energy of the corresponding mean-field state, $\omega = \langle\Psi_{\mathrm{SA}}|H|\Psi_{\mathrm{SA}}\rangle$, and we apply an EFP of degree $d = 32$. 
We compare three initial states with the same $R_{\mathrm{init}}$: the spin-adapted triplet CSF with the largest target overlap
($\gamma^2 = 0.74$), a cooled state with $\sigma = 0.0212$\,Ha
($\gamma^2 = 0.62$), and a random state ($\gamma^2 = 0.04$).
While exact inverse iteration converges within one to two iterations for all three, the QIPI converence varies significantly for the different initial states. 
The spin-adapted state converges quickly to the target state since the engineered initial state is orthogonal to any off-target eigenstates having different particle number or spin symmetry, while the remaining off-target components are sufficiently separated from the target state to be suppressed by the EFP.
The random state first appears to converge rapidly, since most of its weight lies outside $\Delta$ and is suppressed within a few iterations, but then convergence slows down due to the small residual weight on the other eigenstates inside $\Delta$.
The cooled state, despite its larger initial overlap compared to the random state, converges slowly as the energy filter concentrates the off-target weight on a degenerate triplet and a degenerate singlet pair near the target state, for which an EFP of this degree and window only leads to limited suppression.
The comparison shows that a chemically informed state is useful not merely because it can increase the initial target overlap, but because symmetry adaptation removes entire classes of competing eigenstates. 
In particular, when the spin multiplicity and particle number of the desired excited state are known, a corresponding CSF restricts QIPI to the physically relevant symmetry sector. 
This can produce substantially better convergence than energy localization alone (e.g. cooled initial state), even when the latter provides a comparably large initial target overlap.
Symmetry-aware initial states, therefore, benefit from the spectral selectivity, outperforming both energetically pre-filtered and unstructured initial states.

\begin{figure*}[t]
\centering
\includegraphics[width=\linewidth]{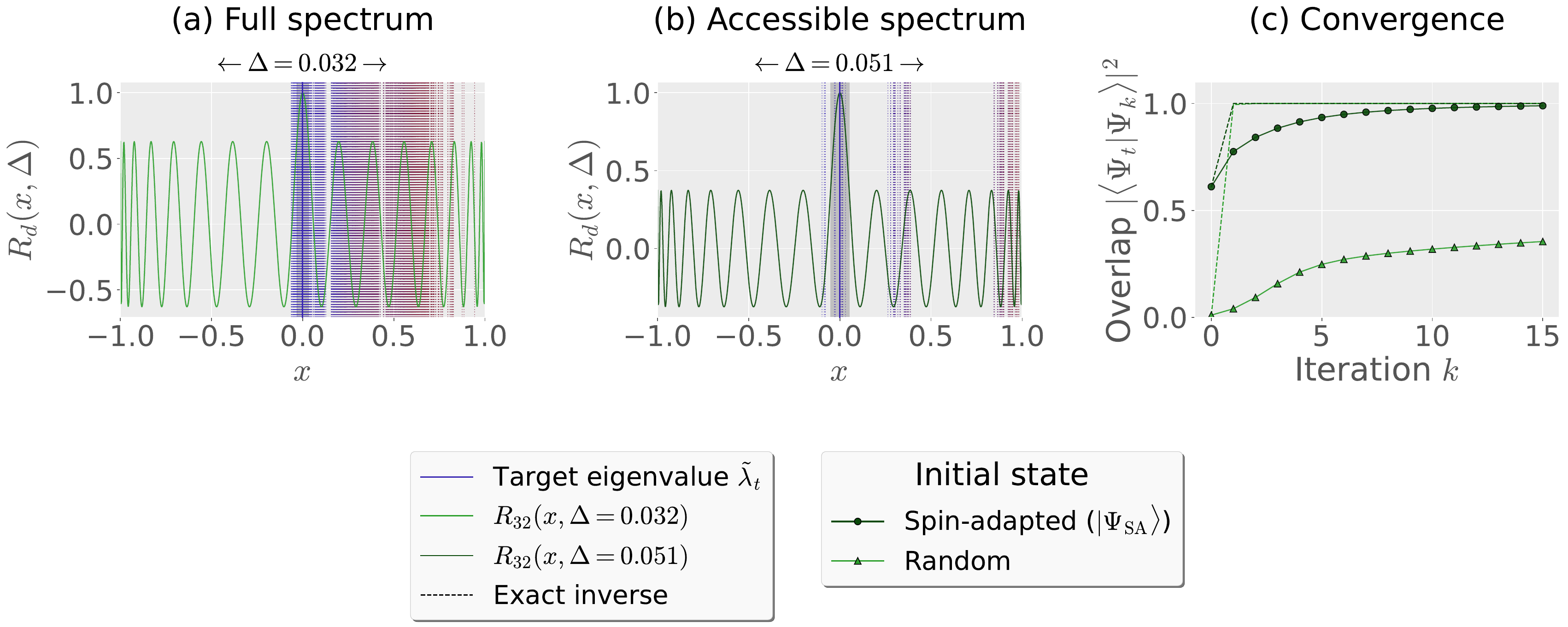}
\caption{Effect of chemically informed symmetry restriction and spectral rescaling on EFP-based QIPI for the 12-qubit Jordan--Wigner Hamiltonian of \ce{LiH} in the STO-3G basis.
(a) Degree-$d=32$ filtering polynomial and the full spectrum, using the Hamiltonian one-norm as the scale factor, resulting in $\Delta=0.032$.
(b) The spectrum accessible from a fixed-particle-number, spin-adapted singlet CSF.
Restricting the problem to the singlet sector permits a tighter scale factor and increases the normalized filtering window to $\Delta=0.051$. 
(c) Target-state overlap during QIPI for a spin-adapted singlet CSF in the accessible singlet sector and a random state in the full Hilbert space with the same $R_{\mathrm{init}}$, and corresponding exact inverse application (dashed curves).
}
\label{fig:jw-accesible}
\end{figure*}

In our second chemically-motivated example, we compare QIPI convergence in the standard setting
with a chemically informed setup, in which the scale $s$ is estimated from classical mean-field calculations and the spectrum is de-cluttered by symmetry.
As shown in Fig.~\ref{fig:jw-accesible}, we consider the full 12-qubit Jordan--Wigner mapped Hamiltonian of strechted \ce{LiH} and its symmetry-reduced spectrum.
As the target we choose a singlet eigenstate and set the shift $\omega$ closest to the target eigenstate in both cases for better comparability.
In the standard setting (Fig.~\ref{fig:jw-accesible}(a)), the full spectrum is rescaled with the one-norm, i. e.  $s = \alpha$, and QIPI is started from a random initial state.
In the chemically informed setting (Fig.~\ref{fig:jw-accesible}(b)), the initial state is the spin-adapted singlet CSF, and $s$ is estimated from the singlet CSF energies by setting it to $s = \max_{\mathrm{SA}} |\langle\Phi_{\mathrm{SA}}|H|\Phi_{\mathrm{SA}}\rangle - \omega|$.
We see in Fig.~\ref{fig:jw-accesible}(c) that in the chemically-informed setup, QIPI converges significantly faster to the target state than in the standard calculation, where the filtering polynomial acts on the complete spectrum, also containing states with different particle numbers and spin quantum numbers.
Although the two initial states have the same $R_{\mathrm{init}}$, the chemically informed construction converges much more rapidly, due  to the removal of symmetry-incompatible sectors, and stronger polynomial compression due to the broader spectral rescaling. 
More generally, incorporating particle number, spin symmetry, and, where available, molecular point-group symmetries into state preparation can reduce the effective spectral density seen by QIPI and consequently lower the polynomial degree or number of iterations required for convergence.

\subsection{Post-selection success probabilities}
\label{subsub: succ-prob}
Since our QIPI convergence analysis involves post-selection after each iteration, we now examine how the choice of initial state affects the cumulative success probability and report representative results for the molecular Hamiltonians considered in this study.
The per-iteration success probability of QIPI is given by $p_k = \|R_d(\tilde{H}, \Delta)\ket{\Psi_{k-1}}\|^2$, with the cumulative success probability amounting to $\Pi_{i = 1}^k p_i$.
As the trial state moves closer and closer to the target state with iterations, the per-iteration success probability is, evidently, not constant across the run.
In early iterations, if we start with an initial state having very small overlap with the target state, 
the filter suppresses the large off-target weight of the trial state, so $p_k$ is smallest at the first iteration, and it then increases monotonically toward $|R_d(\tilde{\lambda}_t,\Delta)|^2 \approx 1$ as the state converges to the target state. 
Representative values for a degree-16 EFP are given in Table~\ref{tab:postselection-probabilities}.
\begin{table}[htbp]
  \centering
  \caption{Representative post-selection probabilities for QIPI using a degree-$16$ EFP and an initial state with a squared target-state overlap of $\gamma^2=0.2$. 
  The reported values are the medians over 25 random trial states iterating to chemical accuracy.}
  \label{tab:postselection-probabilities}
  \begin{tabular}{lcccc}
  \hline
  \hline
  System
  & Iterations $k$
  & First-step $p_1$
  & Final-step $p_k$
  & Cumulative $P_k=\prod_{i=1}^{k}p_i$ \\
  \hline
  \ce{H2}   & 1        & 0.19 & 0.19   & 0.19 \\
  \ce{LiH}  & $\sim 7$ & 0.43 & 0.98   & 0.20 \\
  \ce{BeH2} & $\sim 34$ & 0.54 & 0.9996 & 0.20 \\
  \hline
  \end{tabular}
  \end{table}
The product of the per-iteration success probabilities is given by: $\prod_k p_k = \|P(\tilde{H})^k\ket{\Psi_0}\|^2 \approx \gamma^2\, R_d(\tilde{\lambda}_t,\Delta)^{2k}$, and since $R_d(\tilde{\lambda}_t,\Delta) \approx 1$ this is essentially the initial squared overlap $\gamma^2$, independent of the polynomial degree and of how the amplification is carried over iterations. 
The end-to-end post-selection cost is therefore a property of the initial state.
For \ce{H2}, all off-target eigenvalues lie far outside the filter window, so a single application of the EFP suppresses essentially all off-target weight: $p_1 \approx \gamma^2$ and one iteration reaches convergence. 
For \ce{LiH} and \ce{BeH2}, the denser spectra place many eigenvalues inside or near the $\Delta$ window, where the degree-16 filter suppresses them only weakly and therefore we get contribution from surviving off-target elements  in the sucess probability, increasing it above $\gamma$, this contribution must be removed over subsequent iterations, which is why these systems require more iterations.

For the chemically-guided setting shown in Fig.~\ref{fig:initial-state-comparison} we report the success probabilities in Table \ref{tab:lih-postselection-chem}.
\begin{table*}[htbp]
  \centering
  \caption{
  Post-selection probabilities for degree-$32$ EFP-based QIPI
  applied to stretched \ce{LiH} using different initial states. The iteration counts
  correspond to reaching chemical accuracy.}
  \label{tab:lih-postselection-chem}
  \begin{tabular}{lccccc}
  \hline
  \hline
  Initial state
  & $\gamma^2$
  & Iterations $k$
  & First-step $p_1$
  & Final-step $p_k$
  & $P_k=\prod_{i=1}^{k}p_i$ \\
  \hline
  Spin-adapted CSF        & 0.74 & 15  & 0.89 & 0.988 & 0.65 \\
  Cooled state            & 0.62 & 119 & 0.98 & 0.989 & 0.22 \\
  Random state\textsuperscript{a} & 0.04 & 9 & 0.20 & 0.986 & 0.04 \\
  \hline
  \multicolumn{6}{l}{\footnotesize
  \textsuperscript{a} Median values obtained from 25 random initial states.}
  \end{tabular}
  \end{table*}
When $\omega$ is chosen from the mean-field energy of the initial state, the target state no longer sits exactly at the maximum of the filter (here $R_{32}(\tilde\lambda_t,\Delta) = 0.995$), so each iteration costs a factor $R_d(\tilde\lambda_t,\Delta)^2 \approx 0.99$ and the cumulative success probability decreases slowly with $k$. 
Instead, if $\omega$ is chosen arbitrarily close to the target energy, then $R_d(\tilde{\lambda}_t,\Delta) \to 1$, and the cumulative success probability approaches the initial overlap, $\prod_k p_k \to \gamma^2$.
The spin-adapted initial state is therefore the most favourable, as it converges in the fewest iterations and retains most of its initial overlap as cumulative success probability. 
A large per-iteration success probability alone is not indicative of progress, for instance, the cooled state passes the filter with $p_k \approx 0.98$ from the first iteration, because its off-target weight lies inside the filter window, yet it requires $119$ iterations and its cumulative success probability drops to $0.22$.

\section{Conclusions and Outlook}
\label{sec:conclusions}
In this work, we introduced two approaches to prepare excited states using the quantum inverse power iteration algorithm: one based on a Chebyshev decomposition of the Hamiltonian inverse, and one based on filtering polynomials.
For the latter, instead of approximating the powers of the Hamiltonian in terms of Chebyshev or Fourier series, we effectively emulated its action using an eigenstate filtering polynomial and implemented it on a quantum simulator via block encodings and QSVT.
We found that the decomposition-based Cheb-Inv QIPI method suffers from divergence even when a good initial guess is available, as well as poor compression ability.
EFP-based QIPI resolves this issue by virtue of the symmetric nature of the filtering polynomials:
The EFP construction avoids the numerical instabilities near the spectral origin that typically arise when approximating the inverse function and improves the convergence behaviour compared to previous methods.
We performed proof-of-principle applications of QIPI on molecular Hamiltonians of \ce{H2}, \ce{LiH}, and \ce{BeH2}, and demonstrated that it converges in fewer iterations than existing decomposition-based approaches while maintaining the same query complexity as QSVT.
The QSVT-based implementation incurs only a low, constant ancilla overhead, which does not depend on either the polynomial degree $d$, the spectral window $\Delta$, or the target accuracy $\eta$, since QSVT requires only a single additional ancilla to realize a given polynomial transformations.
While block encoding the Hamiltonian through dilation is a convenient choice for proof-of-principle demonstrations of QIPI, it presupposes direct access to the dilation unitary rather than an efficient circuit for it.
Thus, for practical implementations at scale, more sophisticated block encodings should be considered, for instance structured, sparsity-exploiting block encodings realized via LCU\cite{childs2012lcu}, with coefficient loading handled by primitives such as QROAM\cite{berry2019qubitization}.
Since QIPI inherits the block-encoding-agnostic character of QSVT, it can directly benefit from improved block-encoding constructions as they are developed.
For example, Berry \textit{et al.} report a total $\widetilde{\mathcal{O}}(N^{3/2}\alpha)$ T-gate complexity for one of their qubitization-based quantum-chemistry constructions, where $N$ is the number of orbitals and $\alpha$ is the LCU one-norm.\cite{berry2019qubitization}
More radically, one could also forego block encoding $H$ altogether and instead access its spectrum through controlled time-evolutions in the spirit of the QETU approach\cite{dong2022groundstate} for more early fault-tolerant applications.

By analysing the EFP-based QIPI performance under finite-precision decompositions of $Z$-rotations over the Clifford+T gate set, we showed that QIPI remains effective at rotation precisions compatible with fault-tolerant architectures, making it a practical candidate for implementation on future fault-tolerant quantum computers.
We also expect QIPI to be comparatively robust to noise in early/mid fault-tolerant regimes because of its iterative structure.
Starting from a given initial state, residual noise may cause the state at the $k$-th iteration to drift away from the ideal $\ket{\psi_k}$, but convergence may still be recovered at the cost of additional iterations. 
This makes QIPI favorable relative to single-shot polynomial-application methods, which typically require high polynomial degree and large circuit depth. 
We leave associated noise analysis as future work.

We further investigated the practical performance of QIPI using chemically informed initial states. 
Our results show that convergence depends not only on the
initial target-state overlap, but also on the distribution of off-target spectral weight. 
In particular, symmetry-adapted CSFs exclude competing eigenstates from incompatible particle-number and spin sectors, leading to faster convergence than unstructured or energy-localized initial states. 
Restricting the calculation to the accessible symmetry sector also permits more informed choices of the shift $\omega$, scale factor $s$, and filtering window $\Delta$. These results highlight how chemical information can reduce the effective spectral complexity encountered by QIPI and related algorithms.

Overall, QIPI provides a scalable and versatile framework for excited-state preparation: it can target any desired eigenpair, not just low-lying states, and thus serves as a promising primary algorithmic primitive for applications in quantum chemistry and many-body physics that require access to arbitrary excited states.

\section*{Code Availability}
All code used to generate the results presented in this work is publicly available on github at \url{https://github.com/NQCP/NQCP-AA-git-qipi}.

\begin{acknowledgement}
This work is supported by the Novo Nordisk Foundation, Grant number NNF22SA0081175, NNF Quantum Computing Programme, and partly funded by Innovation Fund Denmark (IFD) under File no. 4356-00009B.
Nina Glaser thanks everyone at Riverlane for their generous hospitality, and Bjorn Berntson and the Discover team for many stimulating discussions on QSVT and fault-tolerant quantum computing in general during her visit.
Srushti Patil thanks Maria Gabriela J. Oliveira for valuable discussions on QSVT, Marcel David Fabian for insights on excited states, and August Luplau Gliese for helping with the chemically motivated initial state preparation.
Both authors thank Lorenzo Laneve and Christoph Sünderhauf for making their \texttt{nlft-qsp} package for finding QSP/QSVT angles publicly available.
\end{acknowledgement}

\bibliography{achemso-demo}

\appendix
\section*{APPENDIX}
\subsection*{Quantum Signal Processing}
\label{app:qsp}
Quantum signal processing (QSP) provides an efficient framework to implement a target polynomial or a close polynomial approximation through a short sequence of phase-modulated single-qubit rotations. 
Typically, two primitive rotations are used: the \emph{signal rotation} \(W\), a rotation about the \(x\)-axis by a fixed angle \(\theta\), and the \emph{signal-processing} rotation \(S\), a rotation about the \(z\)-axis with a sequence of phase angles $\vec{\phi}$ chosen to realize the desired polynomial transformation:
Concretely, we can write $W$ as:
\begin{equation}
    W(a) = 
    \begin{bmatrix}
    a & i\sqrt{1 - a^2} \\
    i\sqrt{1 - a^2} & a
    \end{bmatrix},
\end{equation}
which is an $x$ rotation by angle $\theta = -2 \cos ^{-1} a$ for some scalar $a$. Similarly, $S(\phi)$ can be written as a $z$ rotation by angle $-2\phi$. 
\begin{equation}
    S(\phi) = e^{i \phi Z},
\end{equation}
Here, the sequence of phases $\vec{\phi} = \{\phi_j\}_{j=0}^d$ in $S(\phi_l)$ is selected to implement the prescribed polynomial transformation of at most degree $d$.
Using this sequence of phases, we can write the QSP unitary $U_{\vec{\phi}}$ as:
\begin{equation}
U_{\vec{\phi}} = e^{i\phi_{0}Z}\,\prod_{j=1}^{d} W(a)\, e^{i\phi_{k}Z}.
\end{equation}
It is proven that for a given bounded polynomial $P(a)$ such that $P(a) \in \{-1, 1\}$, there exists a set of QSP phase sequence $\vec{\phi}$ such that $P(a) = \bra{0}U_{\vec{\phi}}\ket{0}$. 
Therefore, matrix element $P(a) = \bra{0}U_{\vec{\phi}}\ket{0}$ is a polynomial in $a$, with degree at most the length of the phase sequence $\vec{\phi}$. 
The polynomials that we get through these alternate phase evolutions are the Chebyshev polynomials of the first kind $\mathcal{T}_d(a)$. 

To be precise, the QSP phase sequence produces a matrix that can be expressed as a polynomial function of $a$: 
\begin{equation}
e^{i\phi_{0}Z}\,\prod_{k=1}^{d} W(a)\,e^{i\phi_{k}Z}
=
\begin{bmatrix}
P(a) & i\,Q(a)\sqrt{1-a^{2}}\\
i\,Q^{*}(a)\sqrt{1-a^{2}} & P^{*}(a)
\end{bmatrix},
\label{eq:qsp-poly-form}
\end{equation}
for $a\in[-1,1]$. For any choice of polynomials $P$ and $Q$ in $a$, one can choose $\vec{\phi}$ so that Eq.~\eqref{eq:qsp-poly-form} holds provided the following conditions are met:
\begin{enumerate}\itemsep2pt
\item $\deg P \le d$ and $\deg Q \le d-1$;
\item $P$ has parity $d \bmod 2$ and $Q$ has parity $(d-1)\bmod 2$;
\item $|P(a)|^{2} + (1-a^{2})\,|Q(a)|^{2} = 1$ for all $a\in[-1,1]$.
\end{enumerate}

In practice, we often care about the induced scalar polynomial on a subsystem rather than the full unitary.
If we set $\mathrm{Poly}(a)=\langle 0|U_{\vec{\phi}}|0\rangle = P(a)$, then feasible choices are restricted to those $P$ admitting a companion $Q$ that satisfies the conditions above, which can be limiting;
A common alternative is to define:
\begin{equation}
\mathrm{Poly}(a)=\langle +|U_{\vec{\phi}}|+\rangle
= \operatorname{Re}\!\big[P(a)\big] + i\,\operatorname{Re}\!\big[Q(a)\big]\sqrt{1-a^{2}}.
\end{equation}
With this choice, for parity matching $d \bmod 2$, one can approximate any real polynomial on $[-1,1]$ with $\deg(\mathrm{Poly})\le d$ and $|\mathrm{Poly}(a)|\le 1$ for all $a\in[-1,1]$, by selecting an appropriate phase sequence $\vec{\phi}$. 
The same QSP sequences can be used to polynomially transform all singular values of a matrix $A$.

\clearpage
  \begin{center}
  \textbf{TOC Graphic}\\[1em]
  \includegraphics[width=3.25in]{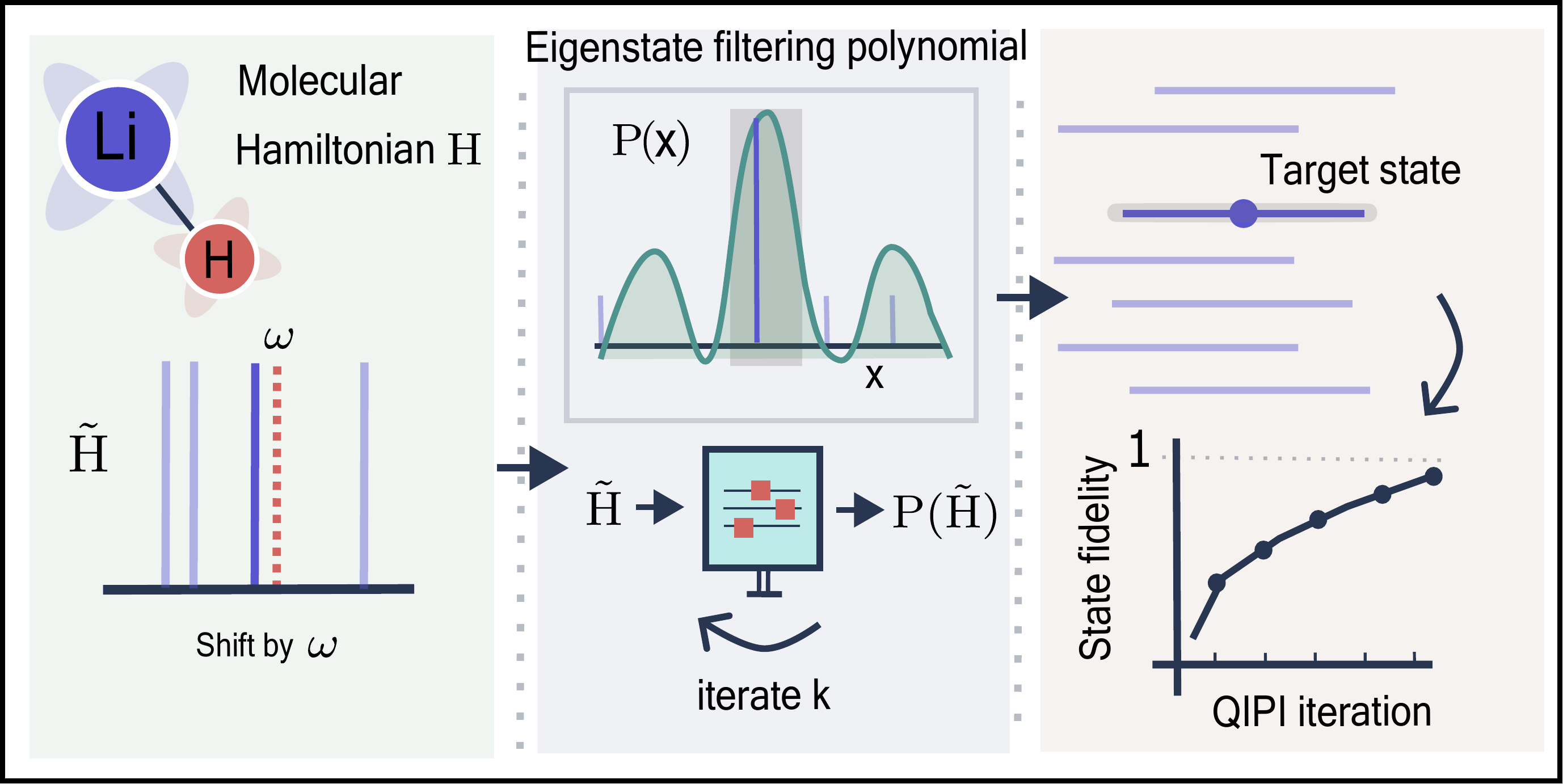}
  \end{center}
  
\end{document}